\documentclass[twoside,11pt]{article}
\usepackage{tikz}
\usetikzlibrary{fit,positioning}

\usepackage{blindtext}

\usepackage{jmlr2e}

\usepackage{amsmath}
\usepackage{amssymb}
\usepackage{bm}
\usepackage{mathtools}
\usepackage{algorithm}
\usepackage{algorithmic}
\usepackage{enumitem}
\usepackage{pgfplots}
\usepackage{tikz}
\usetikzlibrary{calc}

\usepackage{booktabs}
\usepackage{multirow}
\usepackage{changes}

\renewcommand{\thesection}{\arabic{section}.}
\renewcommand{\thesubsection}{\arabic{section}.\arabic{subsection}}

\newtheorem{assumption}{Assmption\assumptionnumber}
\providecommand{\assumptionnumber}{}

\usepackage{lastpage}

\ShortHeadings{Principled Inference in Dense High-Dimensional Linear Models via Local Conditional Sparsity}{Xiong, Chen, Long and Li}
\firstpageno{1}

\begin{document}

\title{Principled Inference in Dense High-Dimensional Linear Models via Local Conditional Sparsity}

\author{\name Wenjun Xiong\thanks{W. Xiong and Y. Chen are co-first authors.} \email wjxiong@gxnu.edu.cn \\
       \addr School of Mathematics and Statistics,\\
Guangxi Normal University\\
Guilin, Guangxi 541004, China
       \AND
       \name Yan Chen$^*$ \email yanchen@amss.ac.cn \\
       \addr State Key Laboratory of Mathematical Sciences\\
Academy of Mathematics and Systems Science\\
Chinese Academy of Sciences,
Beijing 100190, China
\AND
       \name Mingya Long \email mingyalo@umich.edu \\
       \addr 
              Department of Biostatistics, University of Michigan,\\
              Ann Arbor, MI 48109, USA
\AND
       \name Qizhai Li\thanks{Corresponding author} \email liqz@amss.ac.cn \\
       \addr State Key Laboratory of Mathematical Sciences\\
Academy of Mathematics and Systems Science,\\ Chinese Academy of Sciences\\
and University of Chinese Academy of Sciences\\
Beijing 100190, China}

\editor{}

\maketitle

\begin{abstract}
	
	  High-dimensional inference methods often rely on coefficient sparsity, an assumption that can be restrictive when signals are dense but individually weak. In such settings, valid inference may still be possible if the covariates exhibit sparse conditional dependence. Motivated by this observation, we propose Neighborhood-Localized Nested Regression (NLNR), a framework for coordinatewise inference in high-dimensional linear models with potentially dense coefficients. The central idea is to localize inference for a target coefficient to a low-dimensional working regression determined by a Sparse Conditional Neighborhood (SCN) of the target covariate. Specifically, for a given covariate, we estimate its SCN through nodewise $\ell_1$-penalized regression and then fit a regression using only the target covariate and its estimated neighborhood. Under suitable regularity conditions, we establish consistency and asymptotic normality of the resulting estimator. Building on this inferential reduction principle, we further develop a thresholding-based screening procedure with theoretical guarantees and a boosting variant that augments the working model with additional response-relevant covariates to improve finite-sample performance. Extensive simulations and an application to the CCLE dataset demonstrate favorable empirical performance.

\end{abstract}


\section{Introduction}
Recent scientific and technological advances have generated datasets with extremely large numbers of covariates across fields such as medicine, biology, and economics \citep{fan2014challenges, wainwright2019high}.
A defining feature of many such datasets is that the number of covariates $(p_n)$ greatly exceeds the sample size $(n)$. 
This high-dimensional regime poses fundamental challenges for estimation, variable selection, and uncertainty quantification, yet it is increasingly common in modern applications.   
For instance, the Center for Research in Security Prices collected monthly returns for 3{,}680 stocks from 2016 to 2018 to study optimal portfolio management, with only 24 subjects considered \citep{du2023}. 
Another example is the CCLE dataset, which contains $923$ (the sample size) cell lines profiled with $57{,}956$ features, including 56{,}318 gene expression measurements and 1{,}638 mutations \citep{Barretina2012}.

High-dimensional statistical learning has been studied extensively, most often under some form of coefficient sparsity.  Under this paradigm, existing methods can be broadly grouped into three categories. The first category comprises penalized regression methods. Classical examples include the least absolute shrinkage and selection operator (LASSO; \citealp{Tibshirani1996}), smoothly clipped absolute deviation \citep{Fan2001}, the elastic net \citep{Zou2005}, the adaptive LASSO \citep{Zou2006}, the bridge estimator \citep{Huang2008Bridge, Huang2009GroupBridge}, the minimax concave penalty \citep{Zhang2010}, and reciprocal $\ell_1$ regularization \citep{Song2015}, among others. These methods employ sparsity-inducing penalties to reduce effective dimensionality and enhance interpretability \citep{FanLv2010, wainwright2019high}. The second category includes screening-based procedures, such as sure independence screening (SIS), which reduces dimensionality by ranking covariates according to marginal utilities  and then apply penalized regression for refinement \citep{FanLv2010, FanSong2010, LiG2012, LiR2012}.  
The third category focuses on inference for regression parameters, including debiased (or desparsified) LASSO methods \citep{Javanmard2014, VandeGeer2014, Zhang2014} and post-selection inference frameworks \citep{Lee2016, Liang2022, Bao2024}.

A common assumption underlying the aforementioned methods is sparsity of the regression coefficients, i.e., the number of nonzero coefficients $s$ satisfies $s=o\{n/\log(p_n)\}$, where $o(1)$ denotes a term that tends to zero as $n\to\infty$.
This condition is reasonable in many settings; however, it may not hold universally \citep{Cook2012, Bradic2022}. In particular, investigators may encounter settings in which
\[{s}/\{n/\log(p_n)\} \to \infty,\] 
which corresponds to a dense regime in which many covariates have individually weak but collectively non-negligible effects on the response.
Such regimes have been discussed in signal processing \citep{Zhang2008}, genetics \citep{Boyle2017}, and economics \citep{Giannone2022}.
In these settings, inferential procedures whose validity relies critically on coefficient sparsity may break down.

\begin{figure}[t]
	\centering
	\begin{tikzpicture}[scale=1.0, transform shape,
		every node/.style={font=\small},
		cov/.style={
			circle, draw,
			minimum size=6mm,
			inner sep=0pt,
			text height=1.5ex,
			text depth=.25ex
		},
		target/.style={
			circle, draw, thick,
			minimum size=6.5mm,
			inner sep=0pt,
			text height=1.5ex,
			text depth=.25ex
		},
		ynode/.style={
			rectangle, draw, rounded corners=2pt,
			minimum width=8mm, minimum height=6mm
		},
		boxstyle/.style={
			rounded corners=4pt, draw=gray!70, dashed, inner sep=5pt
		}
	]
		
		\node[ynode] (Y1) at (0,0) {$\mathbb Y$};
		
		\node[cov] (X1)   at (-2.3,-1.0) {$\mathbb X_1$};
		\node[cov] (X2)   at (-2.6, 0.0) {$\mathbb X_2$};
		\node[cov] (X3)   at (-2.1, 1.0) {$\mathbb X_3$};
		\node       (Ldot) at (-1.0, 1.65) {$\cdots$};
		
		\node[target] (XjL)   at ( 1.1, 1.65) {$\mathbb X_j$};
		\node         (Rdot1) at ( 1.95, 1.0) {$\cdots$};
		\node[cov, font=\footnotesize] (Xpm1) at ( 2.45, 0.35) {$\mathbb X_{p\!-\!1}$};
		\node[cov]    (Xp)    at ( 2.45,-0.55) {$\mathbb X_p$};
		
		\draw[gray!70]        (X1)   -- (Y1);
		\draw[gray!70]        (X2)   -- (Y1);
		\draw[gray!70]        (X3)   -- (Y1);
		\draw[thick, gray!70] (XjL)  -- (Y1);
		\draw[gray!70]        (Xpm1) -- (Y1);
		\draw[gray!70]        (Xp)   -- (Y1);
		
		\node at (0,-2.35) {Dense global model};
		
		\draw[->, thick] (3.9,0) -- (5.8,0);
		\node[align=center] at (4.85,0.62) {Neighborhood-Induced\\
Inferential Reduction};
		
		\node[target] (Xj) at (8.7,0.15) {$\mathbb X_j$};
		\node[cov] (Xa) at (7.45, 1.45) {$\mathbb X_a$};
		\node[cov] (Xb) at (9.95, 1.45) {$\mathbb X_b$};
		\node[cov] (Xc) at (7.45,-1.15) {$\mathbb X_c$};
		\node[cov] (Xd) at (9.95,-1.15) {$\mathbb X_d$};
		
		\draw[thick] (Xj) -- (Xa);
		\draw[thick] (Xj) -- (Xb);
		\draw[thick] (Xj) -- (Xc);
		\draw[thick] (Xj) -- (Xd);
		
		\node[boxstyle, fit=(Xj)(Xa)(Xb)(Xc)(Xd)] (box) {};
		
		\node[cov] (Xe) at (6.55, 2.2) {$\mathbb X_e$};
		\node[cov] (Xf) at (6.25,-2.65) {$\mathbb X_f$};
		\node[cov] (Xg) at (10.85, 2.2) {$\mathbb X_g$};
		
		\node[ynode] (Y2) at (12.15,0.15) {$\mathbb Y$};
		
		\draw[gray!70] (Xj) -- (Y2);
		\draw[gray!70] (Xa) -- (Y2);
		\draw[gray!70] (Xb) -- (Y2);
		\draw[gray!70] (Xc) -- (Y2);
		\draw[gray!70] (Xd) -- (Y2);
		
		\node[align=center] at (9.45,-3.10)
		{Working model for inference on $\beta_j^*$\\[1mm]
		$\widetilde{\Xi}_j=\{j\}\cup\widetilde{\Omega}_j$};
		
	\end{tikzpicture}
	\caption{Schematic illustration of neighborhood-induced inferential reduction in NLNR. Although the full regression model may be high-dimensional and dense, inference for a target coefficient $\beta_j^*$ is carried out using a low-dimensional working set consisting of the target covariate and its conditional dependence neighborhood. Covariates outside this working set are not included in the localized regression.}
	\label{fig:motivation}
\end{figure}
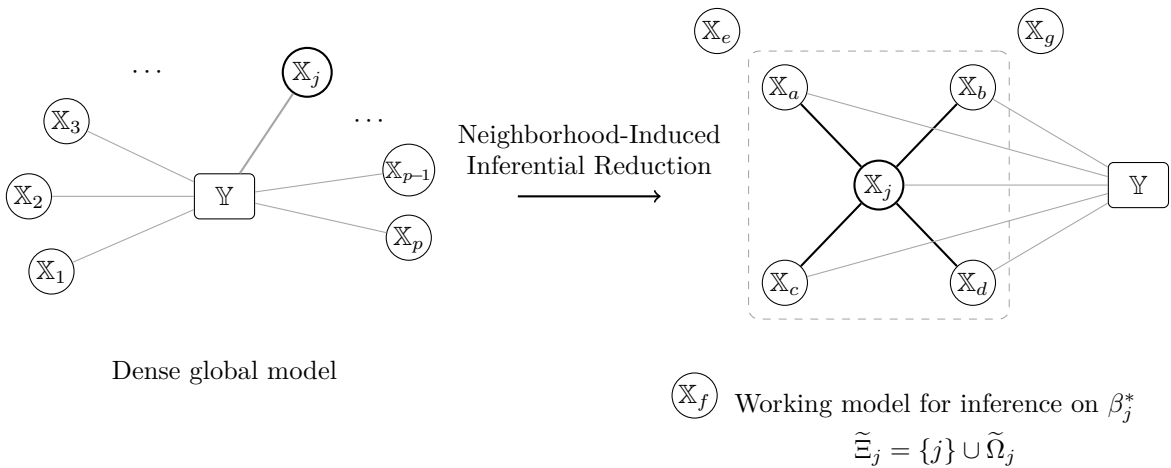

Despite substantial progress in high-dimensional inference, most existing procedures are still developed under coefficient sparsity assumptions. Even when sparsity of the precision matrix is exploited, it is typically used within inferential frameworks whose validity continues to depend on sparse regression coefficients. Consequently, coordinatewise inference in dense-coefficient regimes remains largely unresolved.
A related literature develops methods for learning sparse precision structure, which has become an important ingredient in high-dimensional estimation and inference under coefficient sparsity \citep{VandeGeer2014, Fan2016}. Several methods are available for learning such structure, including nodewise LASSO regression \citep{Meinshausen2006}, $\ell_1$-penalized Gaussian log-likelihood methods \citep{Yuan2007}, and the CLIME estimator based on columnwise constrained $\ell_1$-minimization \citep{cai2011constrained}. While these developments suggest that sparse conditional dependence may provide useful structural information, they do not by themselves yield a direct inferential solution for dense regression vectors. Our contribution is to use this local dependence structure not merely as a technical ingredient, but as the basis for constructing a target-specific low-dimensional working model for inference. Motivated by this perspective, we develop Neighborhood-Localized Nested Regression (NLNR), a framework for coordinatewise inference in dense high-dimensional linear models that leverages Sparse Conditional Neighborhood (SCN) structure to localize the inferential task. In particular, when the covariates exhibit sparse conditional dependence, each target covariate may be associated with only a relatively small SCN, so that inference for the target coefficient can be carried out through a low-dimensional regression involving the target covariate and its neighborhood; see Figure~\ref{fig:motivation} for an illustration.

To formalize the problem, consider the linear regression model
\begin{equation}\label{model}
	\mathbb Y = \beta_0 + \mathbb X_1 \beta_1 + \dots + \mathbb X_{p_n} \beta_{p_n} + \varepsilon,
\end{equation}
where $ \mathbb Y $ is the response variable, $ \mathbb X_1, \dots, \mathbb X_{p_n} $ denote $ p_n $  covariates, and $\varepsilon$ is the error term satisfying ${E}[\varepsilon\mid \mathbb{X}_1, \ldots, \mathbb{X}_{p_n}] = 0$. Let $(\beta_0^*,\beta_1^*,\ldots,\beta_{p_n}^*)^\top$ denote the true parameter. 
Our objective is to conduct coordinatewise inference for $\beta_j^*$ in regimes where $\bm\beta^*$ may be dense but the covariates admit a sparse conditional dependence structure.

Our main contributions are fourfold.

\smallskip
(1) We introduce a target-specific inferential reduction strategy that replaces the sparsity requirement on $\bm\beta^*$ with sparse conditional dependence among covariates. For each $j\in[p_n]=\{1,\ldots,p_n\}$, we estimate the SCN of $\mathbb X_j$ and form a data-adaptive working set
$\widetilde\Xi_j=\{j\}\cup\widetilde\Omega_j$, where $\widetilde\Omega_j$ collects covariates
estimated to be conditionally dependent on $\mathbb X_j$.
This construction reduces the original high-dimensional inference problem to a collection of low-dimensional, covariate-specific working regressions.

\smallskip
(2)
Based on the working set $\widetilde\Xi_j$, we propose a NLNR estimator $\tilde\beta_{j\mid\widetilde\Xi_j}$ for $\beta_j^*$, and establish $\sqrt{n}$-consistency and asymptotically normal inference under growth conditions on the working-model size (Theorem~\ref{theo-semi-012}).
Our theory clarifies how accurate neighborhood recovery controls the remainder terms induced by data-adaptive working models, thereby enabling coordinatewise inference even when \(\bm\beta^*\) is dense.

\smallskip
(3)
Building on the NLNR estimators, we develop thresholding-based variable screening procedures with theoretical guarantees. First, under a beta-min separation condition and a uniform stability condition, we prove exact recovery of the strong-signal set $\bm M^*$ (Theorem~\ref{theo-pscre}). Second, under a milder signal-strength requirement, we prove a sure screening property for a weaker signal set $\bm M_W^*$ (Theorem~\ref{theo-pscre2}).

\smallskip
(4)
To enhance the finite-sample stability, we consider an optional refinement that expands $\widetilde\Xi_j$ by adding a small number of covariates selected via their marginal association with $\mathbb Y$, yielding a boosted NLNR procedure. This refinement is designed to preserve the validity conditions of our theory, requiring only that the expanded working set size remain $o(\sqrt{n})$.
Empirically, it improves the stability of both estimation and screening in finite samples.

The remainder of the paper is organized as follows.
Section~2 introduces the notation and model.
Section~3 presents the NLNR algorithm and an optional boosted NLNR variant to improve finite-sample stability. The framework also remains well-defined and theoretically valid in the fully supervised setting with labeled data only.
Section~4 develops consistent variable screening procedures based on the NLNR and boosted NLNR estimators, with guarantees tailored to dense-coefficient regimes where sparsity-based screening may become less reliable.
Section~5 reports extensive simulation studies, and Section~6 provides a real-data analysis.
Finally, Section~7 concludes the paper.

\section{Notation}
 For a set $M$, let $|M|$ denote its cardinality, i.e., the number of elements in $M$. To simplify notation, the subscript $j$ denotes an index in $[p_n]$ throughout the paper. Let $\mathbb{X} = (\mathbb{X}_1, \dots, \mathbb{X}_{p_n})^\top$ and $\mathbb{X}_{-j} = (\mathbb{X}_1, \dots, \mathbb{X}_{j-1}, \mathbb{X}_{j+1}, \dots, \mathbb{X}_{p_n})^\top$,  where the superscript $\top$ denotes the transpose of a vector or a matrix. 
For an index set $S$, define $\mathbb{X}_S$ as the subvector of $\mathbb{X}$ specified by $S$. For instance, if $S=\{1,3\}$, $\mathbb X_S=(\mathbb X_1,\mathbb X_3)^\top$. We write $a_n \succ b_n$, or equivalently $b_n \prec a_n$, if $b_n / a_n = o(1)$. Define $\bm e_1=(1,0,\ldots,0)^\top$, $\bm e_2=(0,1,0,\ldots,0)^\top$, and let $\bm 1$ denote a vector of ones with dimension clear from the context. Let $\|\cdot\|_q$ denote the standard $\ell_q$ norm of a vector for $q \ge 1$. For a random variable $Z$, define the Orlicz norms
\[
\|Z\|_{\psi_2}
=
\inf\Bigl\{c>0:\ {E}\bigl[\exp(Z^2/c^2)\bigr]\le 2\Bigr\},
\qquad
\|Z\|_{\psi_1}
=
\inf\Bigl\{c>0:\ {E}\bigl[\exp(|Z|/c)\bigr]\le 2\Bigr\}.
\]
We say that $Z$ is sub-Gaussian if $\|Z\|_{\psi_2}<\infty$, and sub-exponential if $\|Z\|_{\psi_1}<\infty$.
For a vector $\bm u$, we write $
\|\bm u\|_\infty = \max_i |u_i|.$
For a matrix, its $\ell_\infty$ norm is defined as the maximum $\ell_1$ norm of its rows.
Denote the ``convergence in probability" and ``convergence in distribution" by ``$\xrightarrow{p}$" and  ``$\xrightarrow{d}$", respectively.
Let $C$ denote a constant, which may take different values in different contexts. Finally, $\mathbb{R}^d$ denotes the $d$-dimensional real space for a positive integer $d$.

We consider data consisting of labeled and possibly additional unlabeled observations.
The labeled data are $\{(\boldsymbol{x}_i, y_i)\}_{i=1}^n$, where $\boldsymbol{x}_i = (x_{i,1}, \dots, x_{i,p_n})^\top \in \mathbb{R}^{p_n}$  and $y_i \in \mathbb{R}$
represent the $i$th observation of $p_n$-dimensional covariates and the response variable, respectively, for $i \in [n]$. The unlabeled observations are $\{\boldsymbol{x}_i\}_{i=n+1}^{n+N}$, where $N \geq 0$, with $N=0$ indicating the absence of unlabeled data.
Denote $\boldsymbol{y} = (y_1, \dots, y_n)^\top$, $\boldsymbol{X} = (\boldsymbol{x}_1, \dots, \boldsymbol{x}_n)^\top$
and $\widetilde{\boldsymbol{X}} = (\boldsymbol{x}_1, \ldots,\bm x_{n+N})^{\top}$. Let $\bm{X}_S$ denote the submatrix of $\bm X$ consisting of the columns indexed by $S$, and $\widetilde{\bm X}_S$ be defined analogously.
Without loss of generality, we assume that all covariates are standardized to have unit variance. Denote the covariance matrix of $\mathbb{X}$ by
$\bm{\Sigma} = {E}[\mathbb{X}\mathbb{X}^\top]$
and assume it is nonsingular. Its inverse
$\bm{\Theta} = \bm{\Sigma}^{-1}$
is the precision matrix. The vector
$\bm{\eta}^*_{-j} = (\eta^*_{-j,1}, \dots, \eta^*_{-j,j-1}, \eta^*_{-j,j+1}, \dots, \eta^*_{-j,p_n})^\top$
is defined as
\[
\bm{\eta}^*_{-j} = \mathop{\arg\min}_{\bm{\eta}_{-j}\in \mathbb{R}^{p_n-1}}
{E}\bigl[(\mathbb{X}_j - \mathbb{X}_{-j}^\top \bm{\eta}_{-j})^2\bigr].
\]
Then $\eta^*_{-j,k}=-\theta_{j,k}/\theta_{j,j}$ for $k\neq j$ \citep{Meinshausen2006,VandeGeer2014}, where $\theta_{j,s}$ denotes the $(j,s)$th entry of $\bm{\Theta}$ for $s\in [p_n]$. Hence, the support of $\bm{\eta}^*_{-j}$ coincides with the set of nonzero off-diagonal entries in the $j$th row of $\bm{\Theta}$. Define
\[
\Omega_j=\{k\in[p_n]\setminus\{j\}:\theta_{j,k}\neq 0\},
~~
\overline{\Omega}_j=\{k\in[p_n]\setminus\{j\}:\theta_{j,k}=0\},
~~
\Xi_j=\{j\}\cup\Omega_j.
\]
Here, $\Omega_j$ is referred to as the SCN of $\mathbb X_j$.

\section{The NLNR Framework}
 For a fixed index $j\in[p_n]$, the inferential target is the population coefficient $\beta_j^*$ in the original regression model, and our goal is to construct confidence intervals and hypothesis tests for this coefficient. The key idea is to replace the original high-dimensional problem by a target-specific low-dimensional working regression determined by the conditional dependence structure of $\mathbb X_j$. For this $j$, the proposed NLNR proceeds in two steps. First, we estimate the SCN $\Omega_j$ by regressing $\mathbb{X}_j$ on $\mathbb{X}_{-j}$ via an $\ell_1$-penalized regression that uses both labeled and unlabeled covariates. Second, we regress $\mathbb{Y}$ on $\mathbb{X}_j$ and the covariates indexed by the estimated SCN, using the labeled sample, and obtain an estimator of $\beta_j^*$ along with a consistent variance estimate. In this way, inference for $\beta_j^*$ is reduced to ordinary least squares on a low-dimensional working model. As shown below, this reduction yields valid coordinatewise inference under the regularity conditions developed for neighborhood recovery and working-model size control. The fully supervised version is obtained by setting $N=0$, in which case only the labeled covariates are used in the SCN estimation step.

\subsection{SCN Recovery}
For each $j\in[p_n]$, we estimate the SCN of $\mathbb X_j$ by fitting an $\ell_1$-penalized regression of $\mathbb X_j$ on the remaining covariates using all available covariate observations, including the unlabeled sample when $N>0$. Specifically, define the penalized estimator
\begin{equation}\label{SS-Nj-lasso}
	\tilde{\bm{{\eta}}}_{-j}
	= \mathop{\arg\min}_{\bm{\eta}_{-j}\in \mathbb{R}^{p_n-1}}
	~
	\frac{1}{2(n+N)} \sum_{i=1}^{n+N} \bigl(x_{i,j} - \boldsymbol{x}_{i,-j}^\top \bm{\eta}_{-j}\bigr)^2 + \lambda_j\|\bm{\eta}_{-j}\|_1,
\end{equation}
where $\tilde{\bm\eta}_{-j} = (\tilde{\eta}_{-j,1}, \dots, \tilde{\eta}_{-j,j-1}, \tilde{\eta}_{-j,j+1}, \dots, \tilde{\eta}_{-j,p_n})^\top$ and $\lambda_j>0$ is a tuning parameter.
We estimate the SCN by
\[
\widetilde{\Omega}_j
=
\Bigl\{k\in[p_n]\setminus\{j\}:\tilde{\eta}_{-j,k}\neq 0\Bigr\}.
\]
We impose the following conditions, which are standard in high-dimensional regression and ensure accurate SCN recovery.

\begin{assumption}[Nodewise regularity]\label{assump:node}

\begin{enumerate}

\item[(a)] 
Let $\bm{\Sigma}={E}(\mathbb{X}\mathbb{X}^{\top})$ and suppose that $\lambda_{\min}(\bm{\Sigma})\ge c_{\Sigma}>0$.
Assume further that there exists a constant $\kappa>0$ such that the standardized design vector $\bm{\Sigma}^{-1/2}\mathbb{X}$ is sub-Gaussian, namely,
\[
\sup_{\|u\|_2=1}\bigl\|u^{\top}\bm{\Sigma}^{-1/2}\mathbb{X}\bigr\|_{\psi_2}\le \kappa.
\]
Moreover, the regression error $\varepsilon$ in~\eqref{model} satisfies ${E}(\varepsilon\mid \mathbb{X})=0$ and is sub-Gaussian with $\|\varepsilon\|_{\psi_2}\le \kappa_{\varepsilon}$. $\operatorname{var}(\mathbb{Y})$ is bounded, that is $\operatorname{var}(\mathbb{Y})\le \bar\sigma_Y^2$ for some constant $\bar\sigma_Y^2$.

\item[(b)] Assume that $\| \{\widetilde{\boldsymbol X}_{\Xi_j}^\top \widetilde{\boldsymbol X}_{\Xi_j} / (n+N)\}^{-1}\|_\infty =o\big(  (n+N)^{1/2 - \gamma-\tilde{\alpha}} \log^{1/2} (n+N)\big)$  and
\[
\big\|
\widetilde{\boldsymbol{X}}_{\overline{\Omega}_j}^{\top}\widetilde{\boldsymbol{X}}_{\Omega_j}
\big(\widetilde{\boldsymbol{X}}_{\Omega_j}^{\top}\widetilde{\boldsymbol{X}}_{\Omega_j}\big)^{-1}
\big\|_{\infty}
\le C<1,
\]
where $\gamma\in(0,1/2)$ and $\tilde{\alpha}\in(0,1/2-\gamma)$.

\item[(c)] Assume that $\min_{s\in\Omega_j} |\eta^*_{-j,s}|
\succ
(n+N)^{-\gamma}\log(n+N),$ $\log (p_n) = o\big((n+N)^{1 - 2\alpha} \log(n+N) \big)$, $ |\Omega_j| =  o(n^{1/2}) $, and
	$\max_{k\neq j} \| \widetilde{\boldsymbol X}_{\{k\}} \|_\infty = o\big( (n+N)^{\alpha} /\log^{1/2} (n+N) \big)$,
where $\alpha\in(1/2-\tilde{\alpha},1/2)$.
Finally, let
$\lambda_j=C_{\lambda,j}(n+N)^{-1/2+\tilde{\alpha}}\log(n+N)$ for a sufficiently large constant $C_{\lambda,j}>0$.

\end{enumerate}
\end{assumption}

\begin{remark}\label{rem:cond_discussion}
\noindent
Assumption~\ref{assump:node}(a)  imposes mild nondegeneracy and tail assumptions that are standard in high-dimensional regression analysis.
In particular, requiring the standardized design $\bm\Sigma^{-1/2}\mathbb X$ to be sub-Gaussian is  less stringent than the Gaussian design assumption often adopted in high-dimensional regression \citep{Meinshausen2006, Zhang2008, Bickel2009, Liang2022}.
Indeed, if $\mathbb X\sim N(\bm 0,\bm\Sigma)$ with $\bm\Sigma\succ\bm 0$, then $\bm Z=\bm\Sigma^{-1/2}\mathbb X\sim N(\bm 0,\bm I)$, so for any $\|u\|_2=1$ we have $u^\top\bm\Sigma^{-1/2}\mathbb X=u^\top\bm Z\sim N(0,1)$.
Moreover, under the Orlicz norm convention $\|Z\|_{\psi_2}=\inf\{c>0:{E}[\exp(Z^2/c^2)]\le 2\}$, a direct calculation yields $\|N(0,1)\|_{\psi_2}\le 2$, and hence
$
\sup_{\|u\|_2=1}\bigl\|u^\top\bm\Sigma^{-1/2}\mathbb X\bigr\|_{\psi_2}\le 2.
$

\smallskip
Assumption~\ref{assump:node}(b)  imposes local $\ell_\infty$-stability of the Gram matrix on the relevant coordinates and an irrepresentability-type restriction.
The first display,
\[
\Bigl\|
\Bigl\{\widetilde{\bm X}_{\Xi_j}^\top \widetilde{\bm X}_{\Xi_j}/(n+N)\Bigr\}^{-1}
\Bigr\|_\infty
=o\Bigl((n+N)^{1/2-\gamma-\tilde\alpha}\log^{1/2} (n+N)\Bigr),
\]
is an $\ell_\infty$-stability (coordinatewise conditioning) requirement for the restricted Gram matrix on $\Xi_j$.
Under the Gaussian design assumption with $\lambda_{\min}(\bm\Sigma_{\Xi_j})\ge c>0$, it holds under mild growth of the submodel size; for instance, it follows when
$|\Omega_j|=o\bigl((n+N)^{1-2\gamma-2\tilde\alpha}\log(n+N)\bigr)$, since
$\bigl\|\{(n+N)^{-1}\widetilde{\bm X}_{\Xi_j}^\top \widetilde{\bm X}_{\Xi_j}\}^{-1}\bigr\|_\infty
=O(\sqrt{|\Omega_j|+1})$. 
The second inequality in (b) is an irrepresentability-type restriction: it bounds the $\ell_1$-norm of the regression coefficients obtained by projecting each ``inactive'' coordinate in $\overline{\Omega}_j$ onto the active set $\Omega_j$.
Such restrictions are well known to be closely tied to support recovery for $\ell_1$-penalized regression (e.g., \citealp{Zhao2006, Wainwright2009, Fan2011}).

\smallskip
Assumption~\ref{assump:node}(c)  specifies detectability and growth-rate requirements, and is analogous to Condition~3 in \citet{Fan2011}. 
The beta-min condition $\min_{s\in\Omega_j}|\eta^*_{-j,s}|\succ (n+N)^{-\gamma}\log(n+N)$ ensures that the nonzero neighborhood coefficients are separated from zero by a margin that dominates the penalization bias.
The restriction $|\Omega_j|=o(n^{1/2})$ keeps the second-stage OLS dimension small enough to deliver $\sqrt{n}$-rate inference, whereas the
growth conditions on $\log(p_n)$ and $\max_{k\neq j}\|\widetilde{\bm X}_{\{k\}}\|_\infty$ control uniform deviations over coordinates.

\end{remark}

We next record a support-recovery result for the estimated neighborhood $\widetilde{\Omega}_j$, which will be used to justify the low-dimensionality of the subsequent working model. 

\begin{lemma}\label{lem:omega-ss}
Fix $j\in[p_n]$. Under Assumption~\ref{assump:node}, we have
\[
\operatorname{pr}\!\left(\widetilde{\Omega}_j= \Omega_j\right)
\ge
1 - 2  |\Omega_j| e^{-(n+N)^{ 2\tilde{\alpha}} \log (n+N)} - 2(p_n -1- |\Omega_j|) e^{-(n+N)^{1 - 2\alpha} \log(n+N)}.
\]
\end{lemma}

Since $|\Omega_j|\le p_n-1$, Lemma~\ref{lem:omega-ss} yields, for each fixed $j\in[p_n]$,
\begin{equation}\label{eq:omega_fail_fixedj_simplified}
\operatorname{pr}(\widetilde{\Omega}_j\neq \Omega_j)
\le
2p_n \exp\!\left\{-(n+N)^{2\tilde{\alpha}}\log(n+N)\right\}
+
2p_n \exp\!\left\{-(n+N)^{1-2\alpha}\log(n+N)\right\}.
\end{equation}
Moreover, under Assumption~\ref{assump:node}(c), namely $\log p_n=o\bigl((n+N)^{1-2\alpha}\log(n+N)\bigr)$ for some $\alpha\in(1/2-\tilde{\alpha},\,1/2)$, the above bound yields that, for any $C>2$, 
\begin{equation*} 
	\operatorname{pr}(\widetilde{\Omega}_j=\Omega_j)\ge 1-o(p_n^{-C-1}). 
\end{equation*}
 For concreteness, we set $C=3$, so that
\begin{equation}\label{eq:omega_fixedj_rate}
\operatorname{pr}(\widetilde{\Omega}_j=\Omega_j)\ge 1-o(p_n^{-4}).
\end{equation}
We will use this bound repeatedly in what follows.

\subsection{Inference via Neighborhood-induced Inferential Reduction}
Lemma~\ref{lem:omega-ss} implies that the event
$\{\widetilde{\Omega}_j=\Omega_j\}$ occurs with high probability, thereby keeping the subsequent working model
low-dimensional.
Let $\widetilde{\Xi}_j=\{j\}\cup\widetilde{\Omega}_j$.
Using the labeled sample $\{(\boldsymbol{x}_i,y_i)\}_{i=1}^n$, we consider the target-specific reduced working model
\begin{equation}\label{tilde-mo}
	\mathbb{Y} = \beta_{0|\widetilde{\Xi}_j} + \mathbb{X}_j \beta_{j|\widetilde{\Xi}_j} + \mathbb{X}_{\widetilde{\Omega}_j}^\top \bm{\gamma}_{\widetilde{\Omega}_j|\widetilde{\Xi}_j} + \tilde{\varepsilon}_j,
\end{equation}
where $\beta_{0|\widetilde{\Xi}_j}$ denotes the intercept, $\beta_{j|\widetilde{\Xi}_j}$ represents the coefficient of $\mathbb{X}_j$, $\bm{\gamma}_{\widetilde{\Omega}_j|\widetilde{\Xi}_j}$ denotes the coefficients corresponding to $\mathbb{X}_{\widetilde{\Omega}_j}^\top$, and the error term $\tilde{\varepsilon}_j$ satisfies $\mathrm{E}(\tilde{\varepsilon}_j) = 0$.

We estimate the regression coefficients in equation \eqref{tilde-mo} using ordinary least squares (OLS), yielding the estimators $(\tilde{\beta}_{0|\widetilde{\Xi}_j}, \tilde{\beta}_{j|\widetilde{\Xi}_j}, \tilde{\bm{\gamma}}_{\widetilde{\Omega}_j|\widetilde{\Xi}_j})$. Assuming that the covariates are centered, we have $\tilde{\beta}_{0|\widetilde{\Xi}_j} = \hat{\beta}_0 = n^{-1}\sum_{i=1}^n y_i$. The remaining coefficients are obtained by
\begin{equation}\label{estimator2}
	(\tilde{\beta}_{j|\widetilde{\Xi}_j}, \tilde{\bm{\gamma}}_{\widetilde{\Omega}_j|\widetilde{\Xi}_j}) = \mathop{\arg\min}_{\beta_{j|\widetilde{\Xi}_j}, \, \bm{\gamma}_{\widetilde{\Omega}_j|\widetilde{\Xi}_j}} \frac{1}{2n}\big\| \boldsymbol{y} - \hat{\beta}_0\bm 1 - \boldsymbol{X}_{\{j\}} \beta_{j|\widetilde{\Xi}_j} - \boldsymbol{X}_{\widetilde{\Omega}_j} \bm{\gamma}_{\widetilde{\Omega}_j|\widetilde{\Xi}_j} \big\|_2^2.
\end{equation}
 Without loss of generality, we assume that the first column of $\boldsymbol{X}_{\widetilde{\Xi}_j}$ contains the observations of $\mathbb{X}_j$. Noting that $\boldsymbol{X}_{\widetilde{\Xi}_j}^\top \bm 1 = 0$ because the covariates are centered, it follows that:
\[
(\tilde{\beta}_{j|\widetilde{\Xi}_j}, \tilde{\bm{\gamma}}_{\widetilde{\Omega}_j|\widetilde{\Xi}_j}^\top)^\top = ( n \widehat{\bm{\Sigma}}_{\widetilde{\Xi}_j} )^{-1} \boldsymbol{X}_{\widetilde{\Xi}_j}^\top \boldsymbol{y},
\]
where $\widehat{\bm{\Sigma}}_{\widetilde{\Xi}_j} = {n}^{-1} \boldsymbol{X}_{\widetilde{\Xi}_j}^\top \boldsymbol{X}_{\widetilde{\Xi}_j}$. We call
\[
\tilde{\beta}_{j\mid\widetilde{\Xi}_j}
=
\bm e_1^{\top}\bigl(n\widehat{\bm{\Sigma}}_{\widetilde{\Xi}_j}\bigr)^{-1}\boldsymbol{X}_{\widetilde{\Xi}_j}^{\top}\boldsymbol{y}
\]
the NLNR estimator of the target coefficient $\beta_j^*$.

For any index set $G_j\subseteq[p_n]$ with $j\in G_j$, let
\[
(\vartheta_{0\mid G_j}^*,\,\bm{\vartheta}_{G_j\mid G_j}^*)
=
\arg\min_{\vartheta_0\in\mathbb R,\ \bm\vartheta\in\mathbb R^{|G_j|}}
{E}\!\left(\mathbb Y-\vartheta_0-\mathbb X_{G_j}^{\top}\bm\vartheta\right)^2,
\]
and define $\varepsilon_{j\mid G_j}
=
\mathbb Y-\vartheta_{0\mid G_j}^*-\mathbb X_{G_j}^{\top}\bm{\vartheta}_{G_j\mid G_j}^*$.
For any such $G_j$, define
\begin{equation}\label{varest-011}
\sigma_{j\mid G_j}^2
=
\frac{1}{n}\,
\bm e_1^{\top}\bm\Sigma_{G_j}^{-1}\,
{E}\!\left[
\mathbb X_{G_j}\mathbb X_{G_j}^{\top}\,
\varepsilon_{j\mid G_j}^2
\right]
\bm\Sigma_{G_j}^{-1}\bm e_1,
\end{equation}
where $\bm e_1$ selects the coordinate of $\mathbb X_j$ within $\mathbb X_{G_j}$ (we order $G_j$ so that $\mathbb X_j$ is its first component).
In general, $(\vartheta_{0\mid G_j}^*,\bm{\vartheta}_{G_j\mid G_j}^*)$ differs from $(\beta_0^*,\bm\beta^*)$ because the NLNR working model omits covariates.

The sample analogue is
\begin{equation}\label{varest-012}
\tilde{\sigma}_{j\mid G_j}^2
=
\frac{1}{n}\,
\bm e_1^{\top}\widehat{\bm\Sigma}_{G_j}^{-1}\,
{E}_n\!\left[
\mathbb X_{G_j}\mathbb X_{G_j}^{\top}\,
\widehat{\varepsilon}_{j\mid G_j}^2
\right]
\widehat{\bm\Sigma}_{G_j}^{-1}\bm e_1,
\end{equation}
where $\widehat{\bm\Sigma}_{G_j}=n^{-1}\boldsymbol X_{G_j}^{\top}\boldsymbol X_{G_j}$, ${E}_n A(\mathbb X, \mathbb Y )= {n}^{-1} \sum_{i=1}^nA(\boldsymbol{x}_i,y_i)$,
$\tilde{\bm\vartheta}_{G_j\mid G_j}=(\boldsymbol X_{G_j}^{\top}\boldsymbol X_{G_j})^{-1}\boldsymbol X_{G_j}^{\top}\boldsymbol y$
is the OLS slope estimator in the working model, and
$\widehat{\varepsilon}_{j\mid G_j}= \mathbb Y-\widehat{\beta}_0-\mathbb X_{G_j}^{\top}\tilde{\bm\vartheta}_{G_j\mid G_j}$.

Therefore, the NLNR estimator $\tilde{\beta}_{j\mid\widetilde{\Xi}_j}$ has asymptotic variance
$\sigma_{j\mid\widetilde{\Xi}_j}^2$.
A consistent variance estimator is obtained by setting $G_j=\widetilde{\Xi}_j$ in \eqref{varest-012},
and the corresponding population functional is given by \eqref{varest-011}. The following theorem establishes the consistency and asymptotic normality of the NLNR estimator.

\begin{theorem}\label{theo-semi-012}
	Under Assumption~\ref{assump:node}, as $n\to\infty$, we have
	$$\tilde{\beta}_{j|\widetilde{\Xi}_j} \xrightarrow{p} \beta_j^*, ~
	\tilde{\sigma}_{j|\widetilde{\Xi}_j}^2 \xrightarrow{p} \sigma_{j|\widetilde{\Xi}_j}^2 \quad \mathrm{and} ~
	n^{1/2} ( \tilde{\beta}_{j|\widetilde{\Xi}_j} - \beta_j^* )/\sigma_{j|\widetilde{\Xi}_j} \xrightarrow{d} {N}\left(0,\, 1 \right).$$
	
\end{theorem}

Inference for $\beta_j^*$ is carried out within the data-adaptive submodel indexed by
$\widetilde{\Xi}_j=\{j\}\cup\widetilde{\Omega}_j$.
By Lemma \ref*{lem:omega-ss}, $|\widetilde{\Omega}_j|$ is of the same order as $|\Omega_j|$ with high probability, so the second-stage regression is low-dimensional even if $\bm \beta^*$ is dense.

\begin{remark}\label{remfull-supervised}
When no unlabeled observations are available ($N=0$), the SCN recovery step
in~\eqref{SS-Nj-lasso} reduces to the nodewise Lasso based on the labeled covariates only.
Consequently, $\tilde{\beta}_{j|\widetilde{\Xi}_j}$ coincides with the fully supervised NLNR estimator. Moreover, all conclusions in Lemma~\ref{lem:omega-ss} and Theorem~\ref{theo-semi-012} remain valid after replacing $(n+N)$ by $n$ throughout,
provided that Assumption~\ref{assump:node} is stated with the same substitution.
The proofs in the appendices do not require $N>0$, and therefore apply verbatim to the fully supervised case $N=0$.
\end{remark}

We also consider joint inference for a finite collection of coefficients, and the corresponding methodology and theory are presented in Appendix A.1.

	\subsection{Boosting NLNR}

	The NLNR estimator $\tilde{\beta}_{j\mid\widetilde{\Xi}_j}$ is obtained by OLS on the working model indexed by
$\widetilde{\Xi}_j=\{j\}\cup\widetilde{\Omega}_j$.
When $\widetilde{\Omega}_j$ recovers the true SCN $\Omega_j$ with high probability (Lemma~\ref{lem:omega-ss}), the second-stage working model remains low-dimensional, which in turn supports valid inference for $\beta_j^*$ under the conditions of Theorem~\ref{theo-semi-012}.

Although NLNR is motivated by the conditional dependence structure of $\mathbb X_j$, it can be beneficial in finite samples to enlarge the working model slightly.
Because $\widetilde{\Xi}_j$ is constructed solely from neighborhood information of $\mathbb X_j$,
it may omit covariates that are weakly related to $\mathbb X_j$ but explain a non-negligible fraction of the variation in $\mathbb Y$.
Omitting such predictors can increase the second-stage residual variation and may inflate the asymptotic variance of
$\tilde{\beta}_{j\mid\widetilde{\Xi}_j}$.
This suggests augmenting the NLNR working model with a small number of response-relevant covariates selected through marginal association scores. The purpose of this augmentation is not to redefine the inferential target, but to improve finite-sample precision by reducing residual variation in the refitted working regression.

Define $\nu_l = |\operatorname{cov}_n(\mathbb{X}_l, \mathbb{Y})|$, where $
\operatorname{cov}_n(\mathbb{X}_l, \mathbb{Y}) = (n-1)^{-1} \sum_{i=1}^n \big(x_{i,l} - {E}_n (\mathbb{X}_l)\big)(y_i - {E}_n \big(\mathbb{Y})\big), $ for $l \in [p_n].$
Given a pre-specified integer $d_j$, let $\tilde{\nu}_{(j,d_j)}$ denote the $d_j$th largest value among
$\{\nu_\ell:\ell\in[p_n]\setminus\widetilde{\Xi}_j\}$, and define
\[
\widetilde{S}_{L,j}
=
\Bigl\{
\ell\in[p_n]\setminus\widetilde{\Xi}_j:\ \nu_\ell\ge \tilde{\nu}_{(j,d_j)}
\Bigr\}.
\]
We then augment $ \widetilde{\Xi}_j $ to form the expanded set $\widetilde{H}_j = \widetilde{\Xi}_j \cup \widetilde{S}_{L,j}.$
We refine the estimator of $\beta_j$ by regressing $\mathbb{Y}$ on
$\mathbb{X}_{\widetilde{H}_j}$, leading to the boosted NLNR
estimator,
\[
\tilde{\beta}_{j\mid\widetilde{H}_j}
=
\bm e_1^{\top}
\bigl(
n\,\widehat{\bm\Sigma}_{\widetilde{H}_j}
\bigr)^{-1}
\boldsymbol X_{\widetilde{H}_j}^{\top}
\boldsymbol y.
\]
Its asymptotic variance $\sigma_{j\mid\widetilde{H}_j}^2$ and the corresponding sample-based estimator
$\tilde{\sigma}_{j\mid\widetilde{H}_j}^2$
are obtained by substituting $G_j=\widetilde{H}_j$ into
\eqref{varest-011} and \eqref{varest-012}, respectively.

To obtain a clean theoretical guarantee, we analyze a sample-splitting version (Algorithm~\ref{alg:refined_inference}) that separates the augmentation step from the refitting step.
Specifically, the marginal association scores are computed on $\mathcal I_1$ by replacing $\operatorname{cov}_n(\cdot,\cdot)$ with
$\operatorname{cov}_{|\mathcal I_1|}(\cdot,\cdot)$, so that $\widetilde S_{L,j}$ is constructed using $\mathcal I_1$ only.
The coefficient and variance estimators are then recomputed on $\mathcal I_2$ using the augmented set $\widetilde H_j$. We denote the refitted coefficient by $\tilde{\beta}_{j\mid\widetilde H_j,\mathcal I_2}$.

Let $\boldsymbol X_{\mathcal I_2,\widetilde H_j}$ denote the submatrix of $\boldsymbol X$ with rows indexed by $\mathcal I_2$ and columns indexed by $\widetilde H_j$,
and let $\boldsymbol y_{\mathcal I_2}$ be the subvector of $\boldsymbol y$ with indices in $\mathcal I_2$.
Define
\[
\widehat{\bm\Sigma}_{\widetilde H_j,\mathcal I_2}
=
|\mathcal I_2|^{-1}\boldsymbol X_{\mathcal I_2,\widetilde H_j}^{\top}\boldsymbol X_{\mathcal I_2,\widetilde H_j},
\qquad
{E}_{\mathcal I_2}A(\mathbb X,\mathbb Y)
=
|\mathcal I_2|^{-1}\sum_{i\in\mathcal I_2}A(\boldsymbol x_i,y_i),
\]
and let
$
\tilde{\bm\vartheta}_{\widetilde H_j\mid \widetilde H_j,\mathcal I_2}
=
(\boldsymbol X_{\mathcal I_2,\widetilde H_j}^{\top}\boldsymbol X_{\mathcal I_2,\widetilde H_j})^{-1}
\boldsymbol X_{\mathcal I_2,\widetilde H_j}^{\top}\boldsymbol y_{\mathcal I_2}.
$
Then
\[
\tilde{\beta}_{j\mid\widetilde H_j,\mathcal I_2}
=
\bm e_1^{\top}
\bigl(
|\mathcal I_2|\,\widehat{\bm\Sigma}_{\widetilde H_j,\mathcal I_2}
\bigr)^{-1}
\boldsymbol X_{\mathcal I_2,\widetilde H_j}^{\top}
\boldsymbol y_{\mathcal I_2}.
\]
Define the corresponding residual
\[
\widehat{\varepsilon}_{j\mid \widetilde H_j,\mathcal I_2}
=
\mathbb Y-\widehat{\beta}_{0,\mathcal I_2}
-
\mathbb X_{\widetilde H_j}^{\top}\tilde{\bm\vartheta}_{\widetilde H_j\mid \widetilde H_j,\mathcal I_2},
\qquad
\widehat{\beta}_{0,\mathcal I_2}=
|\mathcal I_2|^{-1}\sum_{i\in\mathcal I_2}y_i.
\]
The sample-splitting analogue of \eqref{varest-012} (with $G_j=\widetilde H_j$) is
\begin{equation}\label{varest-012-split}
\tilde{\sigma}_{j\mid\widetilde H_j,\mathcal I_2}^2
=
\frac{1}{|\mathcal I_2|}\,
\bm e_1^{\top}\widehat{\bm\Sigma}_{\widetilde H_j,\mathcal I_2}^{-1}\,
{E}_{\mathcal I_2}\!\left[
\mathbb X_{\widetilde H_j}\mathbb X_{\widetilde H_j}^{\top}\,
\widehat{\varepsilon}_{j\mid \widetilde H_j,\mathcal I_2}^2
\right]
\widehat{\bm\Sigma}_{\widetilde H_j,\mathcal I_2}^{-1}\bm e_1 .
\end{equation}

For the theoretical analysis, we require $d_j=o(n^{1/2})$, so that $|\widetilde{H}_j|=o(n^{1/2})$.
In practice, $d_j$ should therefore be chosen conservatively in a way that remains compatible with this growth condition; for example, one may take
$d_j=\lfloor cn^{1/2-\delta}\rfloor$ for some small constants  $c>0$ and $\delta>0$,
or simply use a small fixed value, to keep the refitted model stable while allowing potential precision gains.
Algorithm~\ref{alg:refined_inference} summarizes the procedure.

\begin{algorithm}[htbp]
\caption{Refined Inference via Boosted NLNR}\label{alg:refined_inference}
\textbf{Input:} Labeled data $(\bm X,\bm y)$, an estimated NLNR working set $\widetilde\Xi_j$ for $\mathbb X_j$, and an integer $d_j\ge 0$.\\
\textbf{Output:} Inference of $\beta_j^*$.

\begin{enumerate}
\item Randomly split the labeled indices into two disjoint subsets $\mathcal I_1$ and $\mathcal I_2$ with
$|\mathcal I_2|=\lfloor n/2\rfloor$ and $|\mathcal I_1|=n-|\mathcal I_2|$.
\item Using $\mathcal I_1$, for each $l\in[p_n]\setminus\widetilde\Xi_j$, compute the marginal association score
$\bigl|\operatorname{cov}_{\mathcal I_1}(\mathbb X_l,\mathbb Y)\bigr|$, and let $\widetilde S_{L,j}$ be the indices of the top $d_j$ covariates.
\item Form the expanded set $\widetilde H_j=\widetilde\Xi_j\cup \widetilde S_{L,j}$.
\item Using $\mathcal I_2$, regress $\mathbb Y$ on $\mathbb X_{\widetilde H_j}$ and conduct inference for the coefficient of $\mathbb X_j$.
\end{enumerate}
\end{algorithm}

\begin{theorem}\label{theo-refine}
Under the assumptions of Theorem~\ref{theo-semi-012} and $d_j=o(n^{1/2})$, as $n\to\infty$, we have
\[
\tilde{\beta}_{j\mid\widetilde{H}_j,\mathcal I_2} \xrightarrow{p} \beta_j^*, \qquad
\tilde{\sigma}_{j\mid\widetilde{H}_j,\mathcal I_2}^2 \xrightarrow{p} \sigma_{j\mid\widetilde{H}_j}^2,
\]
and
$
|\mathcal I_2|^{1/2}\bigl(\tilde{\beta}_{j\mid\widetilde{H}_j,\mathcal I_2}-\beta_j^*\bigr)/\sigma_{j\mid\widetilde{H}_j}
\xrightarrow{d} N\!\left(0,\,1\right).
$	
\end{theorem}

Theorem~\ref{theo-refine} is stated for the sample-splitting version because sample splitting decouples the construction of $\widetilde S_{L,j}$ from the refitting step and thereby simplifies the analysis.
For ease of implementation, one may apply Steps~2--4 using the full labeled sample, yielding $\tilde{\beta}_{j\mid\widetilde H_j}$ and
$\tilde{\sigma}_{j\mid\widetilde H_j}^2$.
This full-sample version is the practical implementation used in computation, whereas the theorem below is formulated for the refitted quantities computed on $\mathcal I_2$, with the normalization $|\mathcal I_2|^{1/2}$.

\section{Variable Selection}

\subsection{Selection procedures}
We next consider variable selection based on the NLNR estimates.
To avoid overly complex models in dense regimes, we impose an upper bound $\bar{s}$ on the selected model size, with $\bar{s}\prec p_n$. We define two signal sets. The first set
\[
\bm{M}^{*}
=
\Bigl\{
j \in [p_n]:
|\beta_{j}^{\ast}|\succ n^{-\gamma-\tilde{\alpha}}(\log p_n \log n)^{1/2}
\Bigr\},
~~ |\bm{M}^{*}|=s_0,
\]
corresponds to strong signals. The second set
\[
\bm{M}_W^{*}
=
\Bigl\{
j \in [p_n]:
|\beta_{j}^{\ast}|\succ n^{-\gamma-\tilde{\alpha}}(\log \bar{s} \log n)^{1/2}
\Bigr\},
~~|\bm{M}_W^{*}|=\tilde{s}_0,
\]
represents a weaker signal set associated with the capped model size $\bar s$. In the sequel, we focus on regimes in which $s_0\le \bar{s}$ and $\tilde{s}_0\le \bar{s}$.

Given thresholds $\boldsymbol{\tau}=\{\tau_{nj}: j\in[p_n]\}$, we consider the generic thresholding rule
\begin{equation}\label{eq_varsel}
	\widetilde{\bm{M}}_{\boldsymbol{\tau}|\widetilde{\Xi}}
	= \left\{j \in [p_n]: \big|\tilde{\beta}_{j|\widetilde{\Xi}_j}\big| \ge \tau_{nj} \right\},
\end{equation}
where $\tilde{\beta}_{j\mid\widetilde{\Xi}_j}$ is the NLNR estimator defined in~\eqref{estimator2}.
Algorithm~\ref{alg:varsel} summarizes the procedure.

\begin{algorithm}[htbp]
\caption{Model selection via NLNR}\label{alg:varsel}

\noindent\textbf{Input:} Data $(\widetilde{\boldsymbol X},\boldsymbol y)$ and thresholds $\boldsymbol\tau=\{\tau_{nj}:j\in[p_n]\}$.\\
\textbf{Output:} Selected set $\widetilde{\bm M}_{\boldsymbol\tau\mid\widetilde\Xi}$.

\begin{enumerate}[leftmargin=*, itemsep=2pt, topsep=2pt]
\item 
For each $j\in[p_n]$:
\begin{enumerate}[leftmargin=2em, itemsep=1pt, topsep=1pt, label=\alph*]
\item Estimate the SCN $\widetilde\Omega_j$ (using $\widetilde{\boldsymbol X}$) and set $\widetilde\Xi_j=\{j\}\cup\widetilde\Omega_j$.
\item Fit the reduced OLS regression of $\boldsymbol y$ on $\boldsymbol X_{\widetilde\Xi_j}$ (labeled sample) and obtain $\tilde\beta_{j\mid\widetilde\Xi_j}$.
\end{enumerate}

\item 
Return
\[
\widetilde{\bm M}_{\boldsymbol\tau\mid\widetilde\Xi}
=
\left\{j\in[p_n]:\bigl|\tilde\beta_{j\mid\widetilde\Xi_j}\bigr|\ge \tau_{nj}\right\}.
\]
\end{enumerate}
\end{algorithm}

\begin{assumption}\label{assump:global}
There exist constants $C_{\mathrm{weak}}>0$ and $c_0>0$ such that
\[
\min\{|\beta_j^*|: j\in \bm{M}^{*}\}
\ \succ \
r_n,
\qquad
\max\{|\beta_j^*|: j\notin \bm{M}^{*}\}
\ \le\
C_{\mathrm{weak}}\,r_n,
\]
where $r_n=n^{-\gamma-\tilde{\alpha}}(\log p_n \log n)^{1/2}$, and, uniformly over $j\in[p_n]$,
\[
\big\|
\big(\boldsymbol{X}_{\Xi_j}^{\mathsf T}\boldsymbol{X}_{\Xi_j}/n\big)^{-1}
\big\|_\infty
\le
c_0\,n^{1/2-\gamma-\tilde{\alpha}}(\log n)^{1/2}.
\]
\end{assumption}


\begin{remark}
\smallskip
Assumption~\ref{assump:global} combines a signal-separation requirement with a uniform $\ell_\infty$-stability condition. Together, these conditions support the thresholding analysis for the NLNR estimates. Although the stability condition is stated on the oracle set $\Xi_j$, Lemma~\ref{lem:omega-ss} ensures that $\widetilde{\Xi}_j=\Xi_j$ with high probability.
The first display specifies a two-level beta-min structure at the screening threshold
$r_n=n^{-\gamma-\tilde{\alpha}}(\log p_n \log n)^{1/2}$:
coefficients in the strong-signal set $\bm M^{*}$ dominate $r_n$, whereas coefficients outside $\bm M^{*}$ are uniformly bounded by a constant multiple of $r_n$. This separation is introduced to facilitate exact recovery of $\bm M^*$ through thresholding.

\smallskip
The second display,
\[
\big\|
\big(\boldsymbol{X}_{\Xi_j}^{\mathsf T}\boldsymbol{X}_{\Xi_j}/n\big)^{-1}
\big\|_\infty
\le
c_0\,n^{1/2-\gamma-\tilde{\alpha}}(\log n)^{1/2},
\]
is a uniform $\ell_\infty$-stability (coordinatewise conditioning) requirement for the restricted Gram matrix on $\Xi_j$.
Under a Gaussian design with $\lambda_{\min}(\bm\Sigma_{\Xi_j,\Xi_j})\ge c>0$, this bound holds under mild submodel growth, for example, it holds when
$|\Omega_j|=o\bigl(n^{1/2-\gamma-\tilde{\alpha}}(\log n)^{1/2}\bigr)$.
\end{remark}

To establish recovery of $\bm M^*$, we require SCN recovery uniformly over $j\in[p_n]$.
By \eqref{eq:omega_fixedj_rate} and a union bound, 
\begin{equation*}
\operatorname{pr}\!\left(\widetilde{\Omega}_j=\Omega_j,~\forall j\in[p_n]\right)
\ge 1-o(p_n^{-3}).
\end{equation*}

The following theorem establishes consistent recovery of the strong-signal set $\bm M^*$.

\begin{theorem}\label{theo-pscre}
Suppose that Assumption~\ref{assump:global} holds, and that for each $j\in[p_n]$, 
Assumption~\ref{assump:node} holds.
Let $\tau_{nj}=C_{\tau,j} r_n$, where $C_{\tau,j}> C_{\mathrm{weak}}+c_0 C_0$ and $C_0>0$ is the constant. 
Then, for all sufficiently large $n$,
\[
\operatorname{pr}\!\left(\widetilde{\bm{M}}_{\boldsymbol{\tau}\mid\widetilde{\Xi}}=\bm M^*\right)
\ge 1-\bigl(2+o(1)\bigr)p_n^{-2}
\ge 1-3p_n^{-2}.
\]
\end{theorem}

Although the theoretical selector in \eqref{eq_varsel} allows coordinate-specific thresholds $\{\tau_{nj}\}$, it is useful to consider simplified implementations.
When all thresholds are taken to be equal, namely $\tau_{nj}\equiv \tau_n$ for all $j$, the rule in \eqref{eq_varsel} reduces to a hard-thresholding selector applied to  $\tilde{\beta}_{j\mid\widetilde{\Xi}_j}$.
This is analogous in spirit to the hard-thresholding form
$\hat{\beta}_j^{\mathrm{hard}} = \hat{\beta}_j \cdot I(|\hat{\beta}_j| > \tau_n)$,
commonly used for variable selection, and is also related to screening-type procedures such as SIS
\citep{FanLv2008}.

Because $\tilde{\beta}_{j\mid\widetilde{\Xi}_j}$ is obtained from data-adaptive working regressions with $\widetilde{\Xi}_j$ varying across $j$, its variability may be heterogeneous across coordinates.
To account for this, we recommend standardizing before thresholding.
Specifically, we set
\[
\tau_{nj}=\tilde{\sigma}_{j\mid\widetilde{\Xi}_j}\,\tau_n,
~~ j\in[p_n],
\]
where $\tilde{\sigma}_{j\mid\widetilde{\Xi}_j}$ is the estimated standard error of $\tilde{\beta}_{j\mid\widetilde{\Xi}_j}$
(see Theorem~\ref{theo-semi-012}).
Equivalently, this applies a universal cutoff $\tau_n$ to the studentized statistics
$n^{1/2}\,|\tilde{\beta}_{j\mid\widetilde{\Xi}_j}|/\tilde{\sigma}_{j\mid\widetilde{\Xi}_j}$.
In our numerical studies, we tune $\tau_n$ by cross-validation (Section~4.2).

Theorem~\ref{theo-semi-012} further yields asymptotic normality of $\tilde{\beta}_{j\mid\widetilde{\Xi}_j}$,
which enables a $p$-value based selection.
For testing $H_0:\beta_j^*=0$, define
\[
\tilde{q}_j
=
2\left\{
1-\Phi\!\left(
\frac{n^{1/2}\,|\tilde{\beta}_{j\mid\widetilde{\Xi}_j}|}{\tilde{\sigma}_{j\mid\widetilde{\Xi}_j}}
\right)
\right\},
\]
where $\Phi(\cdot)$ denotes the standard normal cumulative distribution function. These $p$-values provide an asymptotically motivated measure of evidence against the null and can be used to define a nominal selection rule. Given a nominal level $\alpha_{\tau}\in(0,1)$, define
\begin{equation}\label{eq:p_modelsele}
\widetilde{\bm{M}}_{\mathrm{pv}}
=
\left\{j\in[p_n]: \tilde{q}_j\le \alpha_{\tau}\right\}.
\end{equation}
This rule is equivalent to thresholding the studentized statistics at $z_{1-\alpha_{\tau}/2}$, i.e.,
\[
\frac{n^{1/2}\,|\tilde{\beta}_{j\mid\widetilde{\Xi}_j}|}{\tilde{\sigma}_{j\mid\widetilde{\Xi}_j}}
\ge
z_{1-\alpha_{\tau}/2},
\]
or, equivalently, using $\tau_{nj}=n^{-1/2}\tilde{\sigma}_{j\mid\widetilde{\Xi}_j}\,z_{1-\alpha_{\tau}/2}$.
Hence tuning $\alpha_{\tau}$ is equivalent to tuning the universal cutoff $\tau_n=n^{-1/2}z_{1-\alpha_{\tau}/2}$. We adopt the latter implementation in our simulations.

For the weaker signal set $\bm{M}_W^*$, the signals are often too weak to be reliably distinguished from noise or from coefficients outside the set.
Accordingly, we focus on a \emph{screening} guarantee that retains all indices in $\bm M_W^*$ with high probability.

\begin{assumption}\label{assump:weak-complex}
There exists a constant $\widetilde{C}_{\mathrm{weak}}>0$ such that
\[
\min\bigl\{|\beta_j^*|: j\in \bm{M}_W^{*}\bigr\}
\ \succ\
\tilde r_n,
\]
where $\tilde r_n=n^{-\gamma-\tilde{\alpha}}(\log \bar{s}\,\log n)^{1/2}$, and uniformly over $j\in\bm{M}_W^{*}$,
\[ \big\|
\big(\boldsymbol{X}_{\Xi_j}^{\mathsf T}\boldsymbol{X}_{\Xi_j}/n\big)^{-1}
\big\|_\infty 
\le
c_0\,n^{1/2-\gamma-\tilde{\alpha}}(\log n)^{1/2}.
\]
Moreover, letting $d_{\max,W}=\max_{j\in \bm M_W^*} |\Omega_j|$, we have
\[
|\bm M_W^*|\, d_{\max,W} = o(\bar s^{4}).
\]
\end{assumption}

For the sure screening result, it suffices to control SCN recovery uniformly over $j\in\bm M_W^*$.
By \eqref{eq:omega_fixedj_rate} and a union bound over $j\in\bm M_W^*$, 
\begin{equation}\label{eq:omega_uniform_MW}
\operatorname{pr}\!\left(\widetilde{\Omega}_j=\Omega_j,~\forall j\in \bm M_W^*\right)
\ge 1-o(p_n^{-3}).
\end{equation}
The following theorem establishes the sure screening property for $\bm M_W^*$.

\begin{theorem}\label{theo-pscre2}
  Suppose that Assumption~\ref{assump:weak-complex} holds, and that for each $j\in\bm M_W^*$, 
Assumption~\ref{assump:node} holds.
Let $\tau_{nj}=\widetilde{C}_{\tau,j} \tilde r_n$, where $\widetilde{C}_{\tau,j}>0$ is a constant. Then, for all sufficiently large $n$,
\[\operatorname{pr}\!\left(
\bm{M}_W^{*}\subseteq \widetilde{\bm{M}}_{\boldsymbol{\tau}\mid\widetilde{\Xi}}
\right)
\ge
1
-2\sum_{j\in \bm{M}_W^{*}} (|\Omega_j|+1)\,\bar{s}^{-4}
-o(p_n^{-2}).\]
\end{theorem}

The condition $|\bm M_W^*|\, d_{\max,W} = o(\bar s^{4})$ in Assumption~\ref{assump:weak-complex} is imposed to ensure that the leading error term
$2\bar{s}^{-4}\sum_{j\in \bm{M}_W^{*}} (|\Omega_j|+1)$ vanishes, so that the lower bound in
Theorem~\ref{theo-pscre2} converges to $1$.

Classical sure screening procedures such as SIS \citep{FanLv2008} are based on marginal utility rankings. By contrast, our screening strategy is built on the localized NLNR estimators $\tilde{\beta}_{j|\widetilde{\Xi}_j}$, whose consistency is established in Theorem~\ref{theo-semi-012}. Theorem~\ref{theo-pscre2} shows that the resulting selector
$ \widetilde{\bm{M}}_{\boldsymbol{\tau}|\widetilde{\Xi}} $ achieves the sure screening property under the stated conditions. Moreover, one may consider a boosted version of the selector by augmenting the NLNR working model with $\widetilde{S}_{L,j}$, which may improve finite-sample stability.
The corresponding theoretical guarantees are given in Appendix A.2.

\subsection{$K$-fold Cross Validation for Selecting the Threshold $\tau_n$}
In practice, the threshold $\tau_n$ must be chosen in a data-dependent manner. We adopt $K$-fold cross-validation (CV) as a practical tuning rule and use prediction error as the CV criterion. Specifically, the labeled data are partitioned into $K$ folds, and the following steps are performed for each candidate threshold $\tau_n$. For each split, the model is fitted on $K-1$ folds to estimate the SCNs and the corresponding coefficients, and the set $\widetilde{\bm{M}}_{\boldsymbol{\tau}|\widetilde{\Xi}}$ is then constructed using the candidate threshold $\tau_n$. The prediction error is then computed on the held-out fold.
The cross-validation estimate of prediction error is obtained by averaging
${K^{-1}}\sum_{k=1}^{K}\mathrm{CV}_k(\tau_n)$ across the $K$ folds, where $\mathrm{CV}_k(\tau_n)$ denotes the validation error on the $k$th fold.
The selected threshold $\hat{\tau}_n$ is the value of $\tau_n$ that minimizes this average validation error. In our numerical studies, we use $K=5$.

\section{Simulation Studies}\label{sec_simu}
We compare the empirical performance of the proposed NLNR and boosted NLNR estimators with that of the standard LASSO and debiased LASSO (D-LASSO). Unless otherwise specified, tuning parameters in the estimation and inference procedures are selected by five-fold cross-validation.

\subsection{Empirical Estimate}\label{sec_emp_est}
We consider Model \eqref{model} with $\varepsilon\sim N(0,1)$ and $\mathbb{X}\sim N(\mathbf{0},\bm{\Theta}^{-1})$, where  $\bm\Theta= \mathrm{diag}\big\{\bm{\Theta}_1, \dots, \bm{\Theta}_{q}\big\}$  is a block diagonal matrix.   Each block $\bm{\Theta}_{i_0}$ is a $k\times k$ Toeplitz matrix defined by
\[
\big(\bm{\Theta}_{i_0}\big)_{k_1,k_2} = \lambda_{i_0}\rho^{|k_1-k_2|}, \quad k_1,k_2=1,\dots,k,
\]
where $\lambda_{i_0}$ is chosen so that each component of $\mathbb{X}$ has unit variance,
for $i_0=1,\dots,q$ and $q = p_n / k$.
We set $k = 20$, $n=200$ and $N = 6200$.
We consider $\rho=0.6$  and $p_n\in\{60, 120,\ldots, 1200\}$, or $p_n= 500$ and $\rho \in\{0.1,0.2,\ldots,0.7\}$, where the former setting allows us to evaluate the performance of each method as the dimensionality increases and the latter one is used to assess the robustness of each method under different levels of sparsity.
Let the strong signals $\beta_{j_0} = \beta_{20+j_0} = 3.5 - 0.5 j_0$ for $j_0 =1, \ldots, 5$,  which yields the strong signal set $\boldsymbol{M}^{*} = \{1, \ldots, 5, 21, \ldots, 25\}$.
For $j_1 \in [p_n]\backslash \boldsymbol{M}^{*}$,  $\beta_{j_1} = 0.2\sin(j_1)$.
$R=600$ replicates are conducted to calculate the empirical bias and mean squared error (MSE) of $\widehat{\beta}_{j}$ as
\[
\mathrm{Bias}_{j}=\Bigl|\frac1R\sum_{r=1}^R\widehat{\beta}_{j}^{(r)}-\beta_{j}^{\ast}\Bigr|, ~
\mathrm{MSE}_{j}=\frac1R\sum_{r=1}^R\Bigl|\widehat{\beta}_{j}^{(r)} - \beta_{j}^{\ast}\Bigr|^{2},
\]
where $\widehat{\beta}_j^{(r)}$ is the estimate of $\beta_j$ for the $r$th replicate, $r=1,\ldots,R$.
The empirical coverage probability is
\[
\mathrm{CR}_{j} = \frac1R\sum_{r=1}^{R}I\bigl\{\beta_{j}^{\ast}\in[a_{j}^{(r)},b_{j}^{(r)}]\bigr\},
\]
where $[a_j^{(r)}, b_j^{(r)}]$ is the 95\% confidence interval for $\beta_j$ in the $r$th replicate, $r=1,\cdots,R$.

\begin{figure}[htbp]
	\centering
	\includegraphics[width=1\textwidth]{ 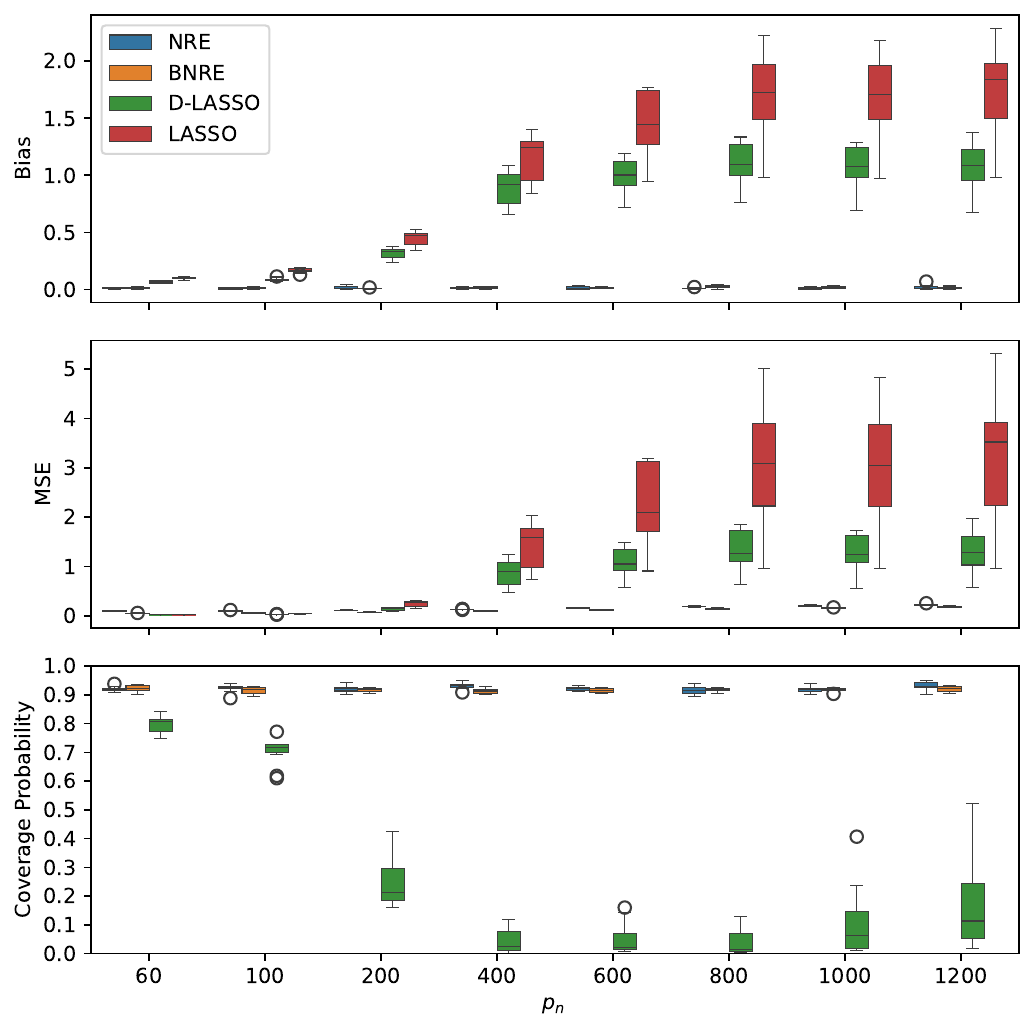}
	\caption{
		The boxplots of empirical biases, MSEs, coverage probabilities of LASSO, D-LASSO, NLNR, and boosted NLNR for 10 strong signals under the setting that $\rho = 0.6$ and different $p_n$ over 600 replicates. Note that there are no empirical coverage probabilities for LASSO.
	}\label{fig_varypn_CrExa1}
\end{figure}

\begin{figure}[htbp]
	\centering
	\includegraphics[width=1\textwidth]{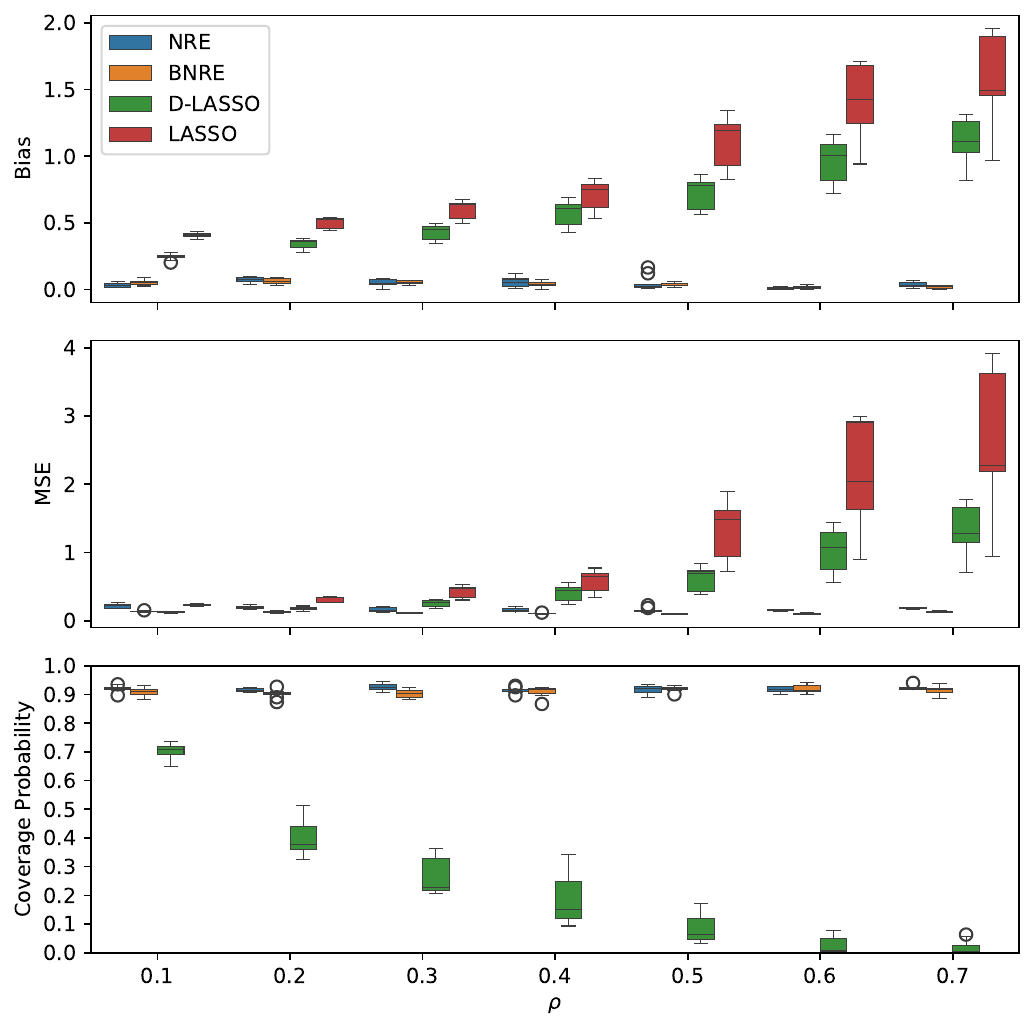}
	\caption{ The boxplots of empirical biases, MSEs, coverage probabilities of LASSO, D-LASSO, NLNR, and boosted NLNR for 10 strong signals under the setting that $p_n = 500$ and  different $\rho$ over 600 replicates. Note that there are no empirical coverage probabilities for LASSO.
	}\label{fig_varyrho_CrExa1}
\end{figure}

Figures~\ref{fig_varypn_CrExa1} and~\ref{fig_varyrho_CrExa1} show the boxplots of the empirical biases, MSEs and coverage rates for the 10 strong signals.
It can be seen that LASSO and D-LASSO have larger biases and MSEs than both the NLNR and the boosted NLNR, especially for large $p_n$ or $\rho$.
For example, when $p_n=1200$ and $\rho=0.6$, the average empirical bias across the strong signals is 0.0173 for the NLNR, 0.0141 for the boosted NLNR, 1.71 for LASSO, and 1.05 for D-LASSO. The corresponding average MSEs for them are 0.220, 0.179, 3.18, and 1.27, respectively.
From Figures 2 and 3, we can also see that both the NLNR and the boosted NLNR achieve higher empirical coverage probabilities than D-LASSO, with the boosted NLNR slightly better than the NLNR.
Note that naive Wald-type confidence intervals for the original LASSO estimator are generally unavailable, since its sampling distribution is nonregular and lacks a tractable limiting form \citep{VandeGeer2014}. The empirical coverage probabilities of the NLNR and the boosted NLNR are close to the nominal 95\% level, whereas D-LASSO exhibits substantially lower coverage probabilities, especially as $p_n$ or $\rho$ increases. For instance, when $p_n=1200$ and $\rho=0.6$, the empirical coverages of the NLNR, the boosted NLNR, and D-LASSO are 93.1\%, 92.1\%, and 18.3\%, respectively. Additional simulation results under varying sparsity levels of the precision matrix $\bm{\Theta}$ are presented in Figure~S1 in Appendix A.3.

\subsection{Variable selection}\label{sec_varsel}

For variable selection, we compare boosted NLNR with D-LASSO.
Two signal structures are considered.
For the sparse structure, we set $\beta_{j_0} = 3 - (j_0 - 1)/4$, $\beta_{20+j_0} = 2 - (j_0 - 1)/4$ for $j_0 = 1,\ldots,5$, and
$\beta_{j_1} = 0$ for $j_1 \in [p_n]\setminus \{1,\ldots,5,20,\ldots,25\}$.
For the dense structure, we set $\beta_{j_0} = 4 - (j_0 - 1)/4$, $\beta_{20+j_0} = 3 - (j_0 - 1)/4$ for $j_0 = 1,\ldots,5$,
and  $\beta_{j_1}=0.2 \sin(j_1)$ for $j_1 \in [p_n]\setminus \{1,\ldots,5,20,\ldots,25\}$. 
For both structures, the strong signal set $\boldsymbol{M}^{*} = \{1, \ldots, 5, 21, \ldots, 25\}$. We adjust the $p$-values in \eqref{eq:p_modelsele} using the Holm-Sidak method \citep{Holm1979}, considering multiple testing, and use it to select the strong signals.
Two thresholds $\alpha_{\tau} = 10^{-3}$ and $10^{-4}$ are considered.
We conduct $R=600$ replicates to compute the proportions $\widetilde{\bm{M}}_{\mathrm{pv}} \supset \bm{M}^{*}$ and $\widetilde{\bm{M}}_{\mathrm{pv}} = \bm{M}^{*}$.
We further assess selection accuracy using the false selection rate (FSR) and the negative selection rate (NSR), defined as
\begin{equation*}
	\mathrm{FSR}= \frac{\sum_{r=1}^R\bigl|\widetilde{\bm{M}}_{\mathrm{pv}}^{(r)} \setminus \boldsymbol{M}^{*}\bigr|}{\sum_{r=1}^R\bigl|\widetilde{\bm{M}}_{\mathrm{pv}}^{(r)}\bigr|}, \quad
	\mathrm{NSR}= \frac{\sum_{r=1}^R\bigl|\boldsymbol{M}^{*}\setminus \widetilde{\bm{M}}_{\mathrm{pv}}^{(r)}\bigr|}{\sum_{r=1}^R\lvert\boldsymbol{M}^{*}\rvert},
\end{equation*}
where $\widetilde{\bm{M}}_{\mathrm{pv}}^{(r)}$ denotes the selected set in the $r$th replicate. Here FSR measures the proportion of selected variables falling outside the target strong-signal set, whereas NSR measures the proportion of strong signals missed by the selection procedure.

\begin{figure}[htbp]
	\centering

\pgfplotstableread[col sep=space]{
method_name pn  include_rate_ci_HS_0.001    exact_rate_ci_HS_0.001  include_rate_ci_HS_0.0001   exact_rate_ci_HS_0.0001
Debiased-Lasso  400 0.976666667 0.885   0.966666667 0.926666667
Debiased-Lasso  600 0.825   0.775   0.765   0.753333333
Debiased-Lasso  800 0.491666667 0.451666667 0.415   0.396666667
Debiased-Lasso  1000    0.19    0.181666667 0.13    0.13
Debiased-Lasso  1200    0.045   0.045   0.033333333 0.033333333
Debiased-Lasso  1400    0.01    0.01    0.005   0.005
Debiased-Lasso  1600    0.011666667 0.011666667 0.005   0.005
}\DlasSparse

\pgfplotstableread[col sep=space]{
method_name pn  include_rate_ci_HS_0.001    exact_rate_ci_HS_0.001  include_rate_ci_HS_0.0001   exact_rate_ci_HS_0.0001
BNR    400 1   0.898333333 1   0.97
BNR    600 1   0.863333333 1   0.943333333
BNR    800 1   0.861666667 1   0.945
BNR    1000    1   0.831666667 1   0.94
BNR    1200    1   0.793333333 1   0.925
BNR    1400    1   0.813333333 1   0.931666667
BNR    1600    1   0.8 1   0.918333333
}\NrSparse

\pgfplotstableread[col sep=space]{
method_name pn  include_rate_ci_HS_0.001    exact_rate_ci_HS_0.001  include_rate_ci_HS_0.0001   exact_rate_ci_HS_0.0001
Debiased-Lasso	400	0.978333333	0.311666667	0.971666667	0.543333333
Debiased-Lasso	600	0.005	0.005	0	0
Debiased-Lasso	800	0	0	0	0
Debiased-Lasso	1000	0	0	0	0
Debiased-Lasso	1200	0	0	0	0
Debiased-Lasso	1400	0	0	0	0
Debiased-Lasso	1600	0	0	0	0
}\DlasDense

\pgfplotstableread[col sep=space]{
method_name pn  include_rate_ci_HS_0.001    exact_rate_ci_HS_0.001  include_rate_ci_HS_0.0001   exact_rate_ci_HS_0.0001
BNR	400	1	0.681666667	1	0.873333333
BNR	600	0.993333333	0.676666667	0.983333333	0.868333333
BNR	800	0.963333333	0.671666667	0.906666667	0.801666667
BNR	1000	0.861666667	0.585	0.746666667	0.648333333
BNR	1200	0.701666667	0.483333333	0.563333333	0.483333333
BNR	1400	0.525	0.335	0.39	0.336666667
BNR	1600	0.358333333	0.23	0.248333333	0.21
}\NrDense

\resizebox{\textwidth}{!}{
  
    \begin{tikzpicture}[remember picture]
      \begin{axis}[
        name=subplot1,
        width=0.4\textwidth,
        xlabel={$p_n$},
        ylabel={Proportion},
        title={Sparse $\bm{\beta}$, $\alpha_{\tau} = 10^{-3}$},
        xmin=400, xmax=1600,
        ymin=0, ymax=1.1,
        xtick={400,800,...,1600},
        ytick={0,0.2,...,1.0},
        grid=both,
        legend to name=commonlegend,
        legend style={
        legend columns=4,
        cells={anchor=west, inner xsep=2pt},
        font=\small,
        /tikz/every even column/.append style={column sep=0.5cm}
        },
        title style={
        yshift=-1.0ex,
        at={(0.5,1.0)},
        anchor=south,
        },
      ]
        \addplot[red, solid, thick, mark=triangle*] table [x=pn, y=include_rate_ci_HS_0.001] {\NrSparse};
        \addlegendentry{BNRE $(\widetilde{\bm{M}}_{\mathrm{pv}} \supset \bm{M}^{*})$}
        
        \addplot[red, densely dashed, thick, mark=diamond*] table [x=pn, y=exact_rate_ci_HS_0.001] {\NrSparse};
        \addlegendentry{BNRE $(\widetilde{\bm{M}}_{\mathrm{pv}} = \bm{M}^{*})$}

        \addplot[blue, solid, thick, mark=*] table [x=pn, y=include_rate_ci_HS_0.001] {\DlasSparse};
        \addlegendentry{D-LASSO $(\widetilde{\bm{M}}_{\mathrm{pv}} \supset \bm{M}^{*})$}

        \addplot[blue, densely dashed, thick, mark=square*] table [x=pn, y=exact_rate_ci_HS_0.001] {\DlasSparse};
        \addlegendentry{D-LASSO $(\widetilde{\bm{M}}_{\mathrm{pv}} = \bm{M}^{*})$}
      \end{axis}

      \begin{axis}[
        name=subplot2,
        at={($(subplot1.east)+(1cm,0)$)},
        anchor=west,
        width=0.4\textwidth,
        xlabel={$p_n$},
        title={Sparse $\bm{\beta}$, $\alpha_{\tau} = 10^{-4}$},
        xmin=400, xmax=1600,
        ymin=0, ymax=1.1,
        xtick={400,800,...,1600},
        ytick={0,0.2,...,1.0},
        grid=both,
        title style={
        yshift=-1.0ex,
        at={(0.5,1.0)},
        anchor=south,
        },
      ]
        \addplot[red, solid, thick, mark=triangle*, forget plot] table [x=pn, y=include_rate_ci_HS_0.0001] {\NrSparse};
        \addplot[blue, solid, thick, mark=*, forget plot] table [x=pn, y=include_rate_ci_HS_0.0001] {\DlasSparse};
        \addplot[red, densely dashed, thick, mark=diamond*, forget plot] table [x=pn, y=exact_rate_ci_HS_0.0001] {\NrSparse};
        \addplot[blue, densely dashed, thick, mark=square*, forget plot] table [x=pn, y=exact_rate_ci_HS_0.0001] {\DlasSparse};
      \end{axis}

      \begin{axis}[
        name=subplot3,
        at={($(subplot2.east)+(1cm,0)$)},
        anchor=west,
        width=0.4\textwidth,
        xlabel={$p_n$},
        title={Dense $\bm{\beta}$, $\alpha_{\tau} = 10^{-3}$},
        xmin=400, xmax=1600,
        ymin=0, ymax=1.1,
        xtick={400,800,...,1600},
        ytick={0,0.2,...,1.0},
        grid=both,
        title style={
        yshift=-1.0ex,
        at={(0.5,1.0)},
        anchor=south,
        },
      ]
        \addplot[red, solid, thick, mark=triangle*, forget plot] table [x=pn, y=include_rate_ci_HS_0.001] {\NrDense};
        \addplot[blue, solid, thick, mark=*, forget plot] table [x=pn, y=include_rate_ci_HS_0.001] {\DlasDense};
        \addplot[red, densely dashed, thick, mark=diamond*, forget plot] table [x=pn, y=exact_rate_ci_HS_0.001] {\NrDense};
        \addplot[blue, densely dashed, thick, mark=square*, forget plot] table [x=pn, y=exact_rate_ci_HS_0.001] {\DlasDense};
      \end{axis}

      \begin{axis}[
        name=subplot4,
        at={($(subplot3.east)+(1cm,0)$)},
        anchor=west,
        width=0.4\textwidth,
        xlabel={$p_n$},
        title={Dense $\bm{\beta}$, $\alpha_{\tau} = 10^{-4}$},
        xmin=400, xmax=1600,
        ymin=0, ymax=1.1,
        xtick={400,800,...,1600},
        ytick={0,0.2,...,1.0},
        grid=both,
        title style={
        yshift=-1.0ex,
        at={(0.5,1.0)},
        anchor=south,
        },
      ]
        \addplot[red, solid, thick, mark=triangle*, forget plot] table [x=pn, y=include_rate_ci_HS_0.0001] {\NrDense};
        \addplot[blue, solid, thick, mark=*, forget plot] table [x=pn, y=include_rate_ci_HS_0.0001] {\DlasDense};
        \addplot[red, densely dashed, thick, mark=diamond*, forget plot] table [x=pn, y=exact_rate_ci_HS_0.0001] {\NrDense};
        \addplot[blue, densely dashed, thick, mark=square*, forget plot] table [x=pn, y=exact_rate_ci_HS_0.0001] {\DlasDense};
      \end{axis}
    \node at ($(subplot1.south west)!0.5!(subplot4.south east)+(0,-2cm)$) {\pgfplotslegendfromname{commonlegend}};
  \end{tikzpicture}
  }
	\caption{Proportions of the events $\widetilde{\bm{M}}_{\mathrm{pv}} \supset \bm{M}^{*}$ and $\widetilde{\bm{M}}_{\mathrm{pv}} = \bm{M}^{*}$ over 600 replicates. Here $\alpha_\tau$ denotes the threshold level.
	}\label{fig_varsel}
\end{figure}
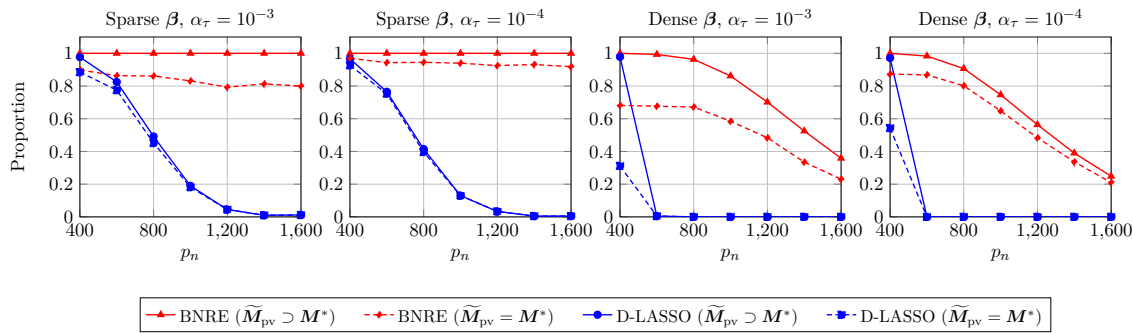

Figure~\ref{fig_varsel} shows the proportions that $\widetilde{\bm{M}}_{\mathrm{pv}} \supset \bm{M}^{*}$ and $\widetilde{\bm{M}}_{\mathrm{pv}} = \bm{M}^{*}$. The left two subfigures correspond to the sparse setting, whereas the right two correspond to the dense setting. Overall, boosted NLNR performs more favorably than D-LASSO in these experiments. Under  the sparse setting, the boosted NLNR could recover $\bm{M}^{*}$ for all the considered value of $p_n$ and the proportion that
$\widetilde{\bm{M}}_{\mathrm{pv}} = \bm{M}^{*}$ exceeds 75\% across all $p_n$ values. In contrast, D-LASSO performs less reliably, with both proportions dropping sharply as $p_n$ increases and approaching zero when $p_n \geq 1200$. For instance, in the second subfigure under the sparse setting with $p_n = 800$ and $\alpha_{\tau} = 10^{-4}$, the proportions that $\widetilde{\bm{M}}_{\mathrm{pv}} \supset \bm{M}^{*}$ and $\widetilde{\bm{M}}_{\mathrm{pv}} = \bm{M}^{*}$
are 100\% and 94.5\% for the boosted NLNR, compared to 41.5\% and 39.7\% for D-LASSO.

\begin{table}[htbp]
	\centering
	\caption{FSR and NSR for boosted NLNR and D-LASSO under sparse and dense $\bm{\beta}$ settings. Here $\alpha_\tau$ denotes the threshold level, $\mathrm{1E2}=1\times10^2$, and the number of replicates is 600. 
	}
	\label{tab_varsel2}
	\resizebox{1.0\textwidth}{!}{
		\begin{tabular}{llllllllll}
			\toprule
			\multirow{2}{*}{Method} &
			\multirow{2}{*}{$p_n$} &
			\multicolumn{4}{c}{Sparse $\bm{\beta}$} &
			\multicolumn{4}{c}{Dense $\bm{\beta}$} \\
			\cmidrule(lr){3-6} \cmidrule(lr){7-10}
			& & \multicolumn{2}{c}{$\alpha_{\tau}=10^{-3}$} & \multicolumn{2}{c}{$\alpha_{\tau}=10^{-4}$} & \multicolumn{2}{c}{$\alpha_{\tau}=10^{-3}$} & \multicolumn{2}{c}{$\alpha_{\tau}=10^{-4}$} \\
			\cmidrule(lr){3-4} \cmidrule(lr){5-6} \cmidrule(lr){7-8} \cmidrule(lr){9-10}
			& & FSR & NSR & FSR & NSR & FSR & NSR & FSR & NSR \\
			\midrule
			boosted NLNR       & 400  & 1.15E-2 & 0.00 & 3.16E-3 & 0.00 & 3.66E-2 & 0.00 & 1.36E-2 & 0.00 \\
			D-LASSO & 400  & 1.04E-2 & 3.17E-3 & 4.67E-3 & 4.50E-3 & 1.13E-1 & 2.33E-3 & 5.96E-2 & 3.00E-3 \\
			\addlinespace
			boosted NLNR       & 600  & 1.53E-2 & 0.00 & 5.63E-3 & 0.00 & 3.79E-2 & 6.67E-4 & 1.24E-2 & 1.67E-3 \\
			D-LASSO & 600  & 9.48E-3 & 2.43E-2 & 3.95E-3 & 3.37E-2 & 1.29E-2 & 5.02E-1 & 7.29E-3 & 5.69E-1 \\
			\addlinespace
			boosted NLNR       & 800  & 1.54E-2 & 0.00 & 5.96E-3 & 0.00 & 3.68E-2 & 4.33E-3 & 1.35E-2 & 1.08E-2 \\
			D-LASSO & 800  & 1.60E-2 & 8.73E-2 & 8.33E-3 & 1.07E-1 & 1.23E-2 & 6.52E-1 & 7.47E-3 & 7.12E-1 \\
			\addlinespace
			boosted NLNR       & 1000 & 2.06E-2 & 0.00 & 6.62E-3 & 0.00 & 3.98E-2 & 1.57E-2 & 1.66E-2 & 3.12E-2 \\
			D-LASSO & 1000 & 1.70E-2 & 1.80E-1 & 1.02E-2 & 2.09E-1 & 1.13E-2 & 7.97E-1 & 6.51E-3 & 8.47E-1 \\
			\addlinespace
			boosted NLNR       & 1200 & 2.34E-2 & 0.00 & 7.94E-3 & 0.00 & 3.90E-2 & 3.90E-2 & 1.63E-2 & 6.35E-2 \\
			D-LASSO & 1200 & 1.56E-2 & 2.54E-1 & 9.13E-3 & 2.77E-1 & 2.44E-2 & 7.67E-1 & 1.31E-2 & 8.12E-1 \\
			\addlinespace
			boosted NLNR       & 1400 & 2.09E-2 & 0.00 & 7.28E-3 & 0.00 & 4.57E-2 & 7.08E-2 & 1.50E-2 & 1.05E-1 \\
			D-LASSO & 1400 & 1.57E-2 & 3.10E-1 & 6.18E-3 & 3.30E-1 & 1.26E-2 & 8.70E-1 & 8.99E-3 & 9.08E-1 \\
			\addlinespace
			boosted NLNR       & 1600 & 2.25E-2 & 0.00 & 8.59E-3 & 0.00 & 4.41E-2 & 1.16E-1 & 1.68E-2 & 1.59E-1 \\
			D-LASSO & 1600 & 2.19E-2 & 3.22E-1 & 1.48E-2 & 3.45E-1 & 1.18E-2 & 8.89E-1 & 6.28E-3 & 9.21E-1 \\
			\bottomrule
		\end{tabular}
	}
\end{table}

Table~\ref{tab_varsel2} reports FSR and NSR for both sparse and dense settings. The boosted NLNR generally achieves lower NSRs than D-LASSO, although the FSR comparison varies somewhat across settings. This pattern indicates that boosted NLNR is more effective at retaining the target strong signals. 
A typical example can be found in the sparse $\bm{\beta}$ setting with $\alpha_{\tau} = 10^{-4}$ and $p_n = 1600$, the boosted NLNR yields an FSR of $8.59 \times 10^{-3}$ and an NSR of 0.00, whereas D-LASSO attains an FSR of $1.48 \times 10^{-2}$ and an NSR of 0.345. Additional variable selection results under both sparse and dense $\bm{\beta}$ configurations are presented in Table~S1 in Appendix A.3.

\section{A Real Application}

We apply the boosted NLNR and D-LASSO to the CCLE data, which is publicly available at \url{https://github.com/alexisbellot/GCIT/tree/master/CCLE\%20Experiments}.
These data include the expression levels of 56,318 genes and 1,638 mutations across 923 cell lines.
Because some genes have replicate measurements, we collapse them by retaining a single value per gene. It yields 54,356 unique genes. Hence, we obtain a total of $p_n = 55{,}994$ covariates, including 54{,}356 unique genes and 1{,}638 mutations.

The drug sensitivity for each cell line is quantified by the area under the dose-response curve, known as the activity area \citep{Barretina2012}, which is used as the response variable.
Among  24 drugs, 21 have the dose-response curve from  $450$ to $460$ cell lines.
For the remaining 3 drugs, the numbers of cell lines are $289$, $381$, and $384$, respectively. To reduce the computational burden associated with the precision-structure estimation step used by D-LASSO, we first apply SIS \citep{FanLv2008} to retain the top 500 features. The boosted NLNR and D-LASSO are then fitted to this prescreened set.

\begin{table}[htbp]
	\caption{Top 1 sensitive gene/mutations selected by boosted NLNR and D-LASSO.
		The values in the parentheses are the width of the 95\% CI length produced by the method.}\label{tab_drug}
	\centering
	\resizebox{\textwidth}{!}{
		\begin{tabular}{lll|lll}
			\toprule
			\textbf{Drug}  & \textbf{boosted NLNR}          & \textbf{D-LASSO}        & \textbf{Drug}  & \textbf{boosted NLNR}          & \textbf{D-LASSO}     \\\midrule
			17-AAG                & NAPG (0.1186)          & NAPG (0.1574)           & PD-0332991            & RB1\_MUT (0.0698)      & LIPJ (0.0972)             \\
			AEW541                & IGLV3-6 (0.0902)       & KCNAB1 (0.1010)         & PF2341066             & SELPLG (0.1089)        & RP11-61D1.2 (0.0891)      \\
			AZD0530               & TRGC2 (0.1705)         & CTSV (0.1500)           & PHA-665752            & TRBV27 (0.1422)        & OR5BA1P (0.0991)          \\
			AZD6244               & RNASE2 (0.1993)        & NRAS\_MUT (0.1221)      & PLX4720               & KRAS (0.0295)          & BRAF.V600E\_MUT (0.1411)  \\
			Erlotinib             & CTD-2117L12.1 (0.0933) & SOX15 (0.1657)          & Paclitaxel            & TUBB3P1 (0.0401)       & ABCB1 (0.1451)            \\
			Irinotecan            & SLFN11 (0.1638)        & SLFN11 (0.1906)         & Panobinostat          & TAF3 (0.1582)          & MYCBP2-AS1 (0.9554)       \\
			L-685458              & ASZ1 (0.1971)          & TRGV5P (0.1416)         & RAF265                & AVP (0.0808)           & GNPTAB (0.1680)           \\
			LBW242                & PDPK1\_MUT (0.0284)    & CXCL1 (0.1191)          & Sorafenib             & KRAS (0.0378)          & RP11-480N24.3 (0.1205)    \\
			Lapatinib             & LSR (0.1999)           & CTB-191K22.5 (0.5311)   & TAE684                & OCSTAMP (0.0638)       & AC003005.2 (0.1082)       \\
			Nilotinib             & SNORA26 (0.0392)       & IL2RG (0.1488)          & TKI258                & LYL1 (0.2145)          & SNX20 (0.2826)            \\
			Nutlin-3              & AC068286.1 (0.2842)    & RP11-255H23.4 (0.1109)  & Topotecan             & SLFN11 (0.1552)        & SLFN11 (0.1730)           \\
			PD-0325901            & RXRG (0.0907)          & SPRY2 (0.1217)          & ZD-6474               & GGTLC1 (0.0932)        & REL\_MUT (0.1037)         \\
			\bottomrule
		\end{tabular}
	}
\end{table}

Table~\ref{tab_drug} presents the most significant genes and mutations selected by the boosted NLNR and D-LASSO for all 24 drugs. The values in parentheses indicate the length of 95\% confidence interval. Below we highlight several representative findings that are supported by prior biological evidence.
(1) For the drugs Irinotecan and Topotecan, both the boosted NLNR and D-LASSO select SLFN11.
Previous studies have shown that SLFN11 is positively associated with response to these two drugs \citep{Barretina2012,Zoppoli2012}.
(2) For  PD-0332991, also known as palbociclib, the boosted NLNR uniquely selects RB1\_MUT (RB1 mutation).
\cite{Condorelli2018Polyclonal} and \cite{Li2022Case} have shown that RB1 mutations emerge under the selective pressure of CDK4/6 inhibitors like palbociclib, and are related to the treatment resistance in metastatic breast cancer patients.
(3) For PLX4720 and Sorafenib, the boosted NLNR identifies KRAS for PLX4720 and Sorafenib, while D-LASSO selects BRAF.V600E\_MUT for PLX4720 and RP11-480N24.3 for Sorafenib.
As reported in \cite{Sullivan2011BRAF}, KRAS mutations can mediate resistance to both drugs via feedback reactivation of MAPK signaling, which provides biological support for the boosted NLNR selection. 
\cite{James2008Discovery} showed that PLX4720 is a highly selective inhibitor of BRAF.V600E\_MUT.
(4) For Paclitaxel, the boosted NLNR identifies TUBB3P1, a pseudogene of TUBB3.
TUBB3P1 may modulate TUBB3 expression by competing for shared miRNA binding \citep{Salmena2011ceRNA,Liu2019DrugResistance}, and the overexpression of TUBB3 is strongly associated with Paclitaxel resistance \citep{Gao2012TUBB3,Mozzetti2005}.
D-LASSO identifies ABCB1, which has also been demonstrated to contribute to paclitaxel resistance \citep{Bo2014paclitaxel}. Taken together, these examples show that several selections made by boosted NLNR are supported by prior biological studies. In addition, boosted NLNR often yields shorter confidence intervals than D-LASSO in this application.

To provide a more detailed illustration, Table~\ref{tab_drugPac} presents the top 10 genes selected for Paclitaxel sensitivity.
The results of other drugs are given in Appendix A.3. 
Both methods identify ABCB1, CDC25A, and BCL2L1; the boosted NLNR further identifies TUBB3P1, TUBB3P2, MCM6, and EEF2.
Each selected gene has prior literature support implicating it in Paclitaxel sensitivity \citep{Mozzetti2005,Salmena2011ceRNA,Gao2012TUBB3,Bo2014paclitaxel,Sun2019CDC25A,Wang2019eEF2,Liu2019DrugResistance,Alalawy2024Key,Song2025Crotonylation}. For completeness, the top-10 sensitive genes identified for the other 23 drugs are reported in Tables~S2–S24 in Appendix A.3.

\begin{table}[htbp]
	\caption{
		Top 10 sensitive genes selected by boosted NLNR and D-LASSO for the drug Paclitaxel.
	}\label{tab_drugPac}
	\centering
		\resizebox{\textwidth}{!}{
			\begin{tabular}{c|cccccccccc}
				\toprule
				\textbf{Rank} & \textbf{1} & \textbf{2} & \textbf{3} & \textbf{4} & \textbf{5} & \textbf{6} & \textbf{7} & \textbf{8} & \textbf{9} & \textbf{10} \\
				\midrule
				boosted NLNR     & TUBB3P1 & ABCB1 & EEF2 & CDC25A & MCM6 & BCL2L1 & SMARCC1 & TUBB3P2 & LMNB1 & NUP188 \\
				D-LASSO  & ABCB1 & BCL2L1 & CDC25A & QARS & ELOVL1 & GIPC1 & PCDHGA2 & CBFB & SLC25A16 & MAGED2 \\
				\bottomrule
			\end{tabular}
		}
\end{table}

\section{Conclusion}
This paper proposes NLNR for coordinatewise inference in high-dimensional linear models with potentially dense regression coefficients. In contrast to methods that rely on coefficient sparsity, NLNR exploits sparse conditional neighborhood structure among the covariates to construct target-specific low-dimensional working regressions. Under the regularity conditions developed in the paper, the resulting estimator is consistent and asymptotically normal for each target coefficient.

We further develop variable selection procedures based on the NLNR estimators. For strong signals, we establish exact recovery of the target strong-signal set, and for weaker signals, we establish a sure screening property. These results are obtained under conditions on neighborhood recovery, local stability, and working-model size, and they allow dense regression vectors provided that the conditional dependence structure remains sufficiently sparse.

We also consider an optional boosted version of NLNR to improve finite-sample precision. In addition, we extend the framework to joint inference on a fixed collection of coefficients, with details of the joint-inference methodology and theoretical results provided in Appendix A.1.

\appendix
\renewcommand{\thesection}{\Alph{section}}
\renewcommand{\thesubsection}{\thesection.\arabic{subsection}}

\section{Additional Methodological Details and Experiments}
\label{sec:remark1_aux}

Appendix~A is organized as follows.
Section~A.1 presents an extension to joint inference for a subset of coefficients.
Sections~A.2 presents further methodological details on variable selection based on the boosted NLNR.
Sections~A.3 reports the supplementary simulation results, including the additional settings and figures that complement numerical studies in the main paper.

\subsection{Joint inference for multiple coefficients}
\renewcommand{\thetheorem}{S\arabic{theorem}}

\noindent
We next extend NLNR to joint inference for a fixed collection of coefficients.
Fix a subset $T\subset[p_n]$ with $|T|$ not depending on $n$, and consider testing the joint null $\bm{\beta}_T=\bm 0$. Define \(\Xi_T = \cup_{j\in T}\Xi_j\). We assume that the joint working set satisfies \(|\Xi_T|=o\bigl(n^{1/2}\bigr).\)  When \(\Xi_T\) is unknown, we estimate it by \(\widetilde{\Xi}_T = \cup_{j\in T}\widetilde{\Xi}_j\). Analogously to the univariate case, we work with the following target-specific working regression:
\[
  \mathbb{Y} =\beta_{0|\widetilde{\Xi}_T}+ \mathbb{X} _{\widetilde{\Xi}_T}^\top \bm{\gamma}_{\widetilde{\Xi}_T} +  \varepsilon_T,
\]
where \(\beta_{0|\widetilde{\Xi}_T}\) denotes the intercept, \(\bm{\gamma}_{\widetilde{\Xi}_T}\) denotes the coefficients of \(\mathbb{X}_{\widetilde{\Xi}_T}\), and \(\varepsilon_T\) meets \({E}[\varepsilon_T] = 0\).
The OLS estimator \(\tilde{\bm{\gamma}}_{\widetilde{\Xi}_T|\widetilde{\Xi}_T}=\bigl(n\widehat{\bm{\Sigma}}_{\widetilde{\Xi}_T}\bigr)^{-1}\boldsymbol{X}_{\widetilde{\Xi}_T}^\top\boldsymbol{y}\) has covariance matrix
\[
\bm{A}_{\widetilde{\Xi}_T} = \frac{1}{n}\bigl(\bm{\Sigma}_{\widetilde{\Xi}_T}\bigr)^{-1}\,{E}\bigl[\mathbb{X} _{\widetilde{\Xi}_T}\mathbb{X} _{\widetilde{\Xi}_T}^\top\,(\mathbb{Y}-\mathbb{X} _{\widetilde{\Xi}_T}^\top\bm{\gamma}_{\widetilde{\Xi}_T}^*)^2\bigr]\,\bigl(\bm{\Sigma}_{\widetilde{\Xi}_T}\bigr)^{-1},
\]
where \(\bm{\gamma}_{\widetilde{\Xi}_T|\widetilde{\Xi}_T}^*=\mathop{\hbox{argmin}}_{\bm{\gamma}_{\widetilde{\Xi}_T}}{E}\bigl[(\mathbb{Y}-\mathbb{X} _{\widetilde{\Xi}_T}^\top\bm{\gamma}_{\widetilde{\Xi}_T})^2\bigr]\). Thus, \(\bm{\gamma}_{\widetilde{\Xi}_T|\widetilde{\Xi}_T}^*\) denotes the population least-squares projection coefficient under the joint working set \(\widetilde{\Xi}_T\). Assuming that the coordinates in $\widetilde{\Xi}_T$ are ordered so that the components of $T$ appear first,
the covariance matrix of $\tilde{\bm{\gamma}}_{T|\widetilde{\Xi}_T}$ can be written as
\[\bm{A}_{T|\widetilde{\Xi}_T}=\hbox{diag}\big(I_{|T|\times |T|},0_{|T|\times (|\widetilde{\Xi}_T|-|T|)}\big)\bm{A}_{\widetilde{\Xi}_T}\hbox{diag}\big(I_{|T|\times |T|},0_{|T|\times (|\widetilde{\Xi}_T|-|T|)}\big)^\top,\] where \(I_{|T|\times |T|}\) denotes the \(|T|\times |T|\) identity matrix and \(0_{|T|\times (|\widetilde{\Xi}_T|-|T|)}\) denotes the \(|T|\times (|\widetilde{\Xi}_T|-|T|)\) zero matrix.
If \(\bm{A}_{T|\widetilde{\Xi}_T}\) is unknown, it can be estimated by
\[\widetilde{\bm{A}}_{T|\widetilde{\Xi}_T}=\hbox{diag}\big(I_{|T|\times |T|},0_{|T|\times (|\widetilde{\Xi}_T|-|T|)}\big)\widetilde{\bm{A}}_{\widetilde{\Xi}_T}\hbox{diag}\big(I_{|T|\times |T|},0_{|T|\times (|\widetilde{\Xi}_T|-|T|)}\big)^{\top},\] where
\(
\widetilde{\bm{A}}_{\widetilde{\Xi}_T}
\;=\;\frac{1}{n}
\bigl(\widehat{\bm{\Sigma}}_{\widetilde{\Xi}_T}\bigr)^{-1}
{E}_n\bigl[\mathbb{X} _{\widetilde{\Xi}_T}\mathbb{X} _{\widetilde{\Xi}_T}^\top\,(\mathbb{Y} - \mathbb{X} _{\widetilde{\Xi}_T}^\top\,\bm{\gamma}^*_{\widetilde{\Xi}_T})^2\bigr]\,
\bigl(\widehat{\bm{\Sigma}}_{\widetilde{\Xi}_T}\bigr)^{-1}.
\)

\begin{theorem}\label{theo-joint}
Suppose that, for each $j \in T$, Assumption~\ref{assump:node} holds, and that $|T \cup \Xi_T| = o(n^{1/2})$. Then,
$\tilde{\bm{\gamma}}_{T|\widetilde{\Xi}_T} \xrightarrow{p} \beta_T^*, ~
\widetilde{\bm{A}}_{T|\widetilde{\Xi}_T} \xrightarrow{p} \bm{A}_{T|\widetilde{\Xi}_T}, ~ \text{and}~
n^{1/2}\bigl(\tilde{\bm{\gamma}}_{T|\widetilde{\Xi}_T} - \bm{\beta}_T^*\bigr)\,\xrightarrow{d}\,{N}\bigl(\bm{0},\,\bm{A}_{T|\widetilde{\Xi}_T}\bigr),$ as \( n \to \infty \).
  \end{theorem}

This theorem implies that $\tilde{\bm{\gamma}}_{T|\widetilde{\Xi}_T}$ follows an asymptotic multivariate normal distribution, which enables the construction of rejection regions for testing $\bm{\beta}_T=\bm{0}$. Specifically, the null hypothesis is rejected when
  \[
  n\,\tilde{\bm{\gamma}}_{T|\widetilde{\Xi}_T}^\top\,\widetilde{\bm{A}}_{T|\widetilde{\Xi}_T}^{-1}\,\tilde{\bm{\gamma}}_{T|\widetilde{\Xi}_T}
  \;\ge\;\chi^2_{1-\alpha}(\lvert T\rvert),
  \]
  where \(\chi^2_{1-\alpha}(\lvert T\rvert)\) is the \((1-\alpha)\)-quantile of the chi-square distribution with \(\lvert T\rvert\) degrees of freedom. The corresponding \((1-\alpha)\) confidence region for \(\bm{\beta}_T\) is
  \[
  \Bigl\{\bm{\beta}_T\colon\;n\,(\tilde{\bm{\gamma}}_{T|\widetilde{\Xi}_T} - \bm{\beta}_T)^\top\,\widetilde{\bm{A}}^{-1}_{T|\widetilde{\Xi}_T}\,
  (\tilde{\bm{\gamma}}_{T|\widetilde{\Xi}_T} - \bm{\beta}_T)\;\le\;\chi^2_{1-\alpha}(\lvert T\rvert)\Bigr\}.
  \]
  Consequently, \(\tilde{\bm{\gamma}}_{T|\widetilde{\Xi}_T}\) maintains asymptotic properties analogous to Theorem \ref*{theo-semi-012}.

\subsection{Variable Selection via boosted NLNR}
\renewcommand{\thetheorem}{S\arabic{theorem}}
\noindent

Theorem~\ref*{theo-refine} shows that the boosted NLNR estimator is asymptotically valid under the expanded working model. Motivated by this result, we consider a threshold-based variable selection procedure built from 
\(\{\tilde{\beta}_{j\mid\widetilde{H}_j,\mathcal I_2}: j\in[p_n]\}\). Given thresholds $\boldsymbol\tau=\{\tau_{nj}:j\in[p_n]\}$, define the expanded-set selector
\[
\widetilde{\bm M}_{\boldsymbol\tau\mid\widetilde H,\mathcal I_2}
=
\left\{j\in[p_n]:\bigl|\tilde\beta_{j\mid\widetilde H_j,\mathcal I_2}\bigr|\ge \tau_{nj}\right\}.
\]
As in the main text, one may also threshold the studentized statistics by taking
$
\tau_{nj}=\tilde\sigma_{j\mid\widetilde H_j,\mathcal I_2}\tau_n.
$

\vspace{0.1in}

\noindent\textbf{Condition:}
\begin{enumerate}

\item[(Su1)] There exist constants $C_{\mathrm{weak},H}>0$ and $c_{0,H}>0$ such that
\[
\min\{|\beta_j^*|: j\in \bm{M}^{*}\}
\ \succ \
r_n,
\qquad
\max\{|\beta_j^*|: j\notin \bm{M}^{*}\}
\ \le\
C_{\mathrm{weak},H}\,r_n,
\]
where $r_n=n^{-\gamma-\tilde{\alpha}}(\log p_n \log n)^{1/2}$, and, uniformly over $j\in[p_n]$,
\[
\left\|
\Bigl(\boldsymbol{X}_{\mathcal I_2,\widetilde H_j}^{\mathsf T}\boldsymbol{X}_{\mathcal I_2,\widetilde H_j}/|\mathcal I_2|\Bigr)^{-1}
\right\|_\infty
\le
c_{0,H}\,n^{1/2-\gamma-\tilde{\alpha}}(\log n)^{1/2}.
\]

\end{enumerate}

\begin{theorem}\label{theo-pscre-H}
Suppose Assumption~\ref{assump:node} hold for $j\in[p_n]$, and Condition \textup{(Su1)} holds.
Let $\tau_{nj}=C_{\tau,j} r_n$ with $C_{\tau,j}> C_{\mathrm{weak},H}+c_{0,H} C_0$.
Then, for all sufficiently large $n$,
\[
\operatorname{pr}\!\left(\widetilde{\bm{M}}_{\boldsymbol{\tau}\mid\widetilde{H},\mathcal I_2}=\bm M^*\right)
\ge 1-\bigl(2+o(1)\bigr)p_n^{-2}\ge 1-3p_n^{-2}.
\]
\end{theorem}

\medskip
Although the theoretical selector based on $\{\tau_{nj}\}$ allows coordinate-specific thresholds,
it is useful to consider simplified implementations.
In particular, one may take $\tau_{nj}\equiv \tau_n$ for all $j$ to obtain a hard-thresholding selector applied to
the boosted NLNR estimates $\tilde{\beta}_{j\mid\widetilde H_j,\mathcal I_2}$.
As in the main text, we recommend standardizing before thresholding to accommodate heterogeneous variability across coordinates.
Specifically, we may set
\[
\tau_{nj}=\tilde{\sigma}_{j\mid\widetilde H_j,\mathcal I_2}\,\tau_n,
\qquad j\in[p_n].
\]
Equivalently, this applies a universal cutoff $\tau_n$ to the studentized statistics
$|\mathcal I_2|^{1/2}\,|\tilde{\beta}_{j\mid\widetilde H_j,\mathcal I_2}|/\tilde{\sigma}_{j\mid\widetilde H_j,\mathcal I_2}$.
A $p$-value-based alternative can also be used, analogously to \eqref{eq:p_modelsele} in Section~4.1, based on the asymptotic normality of the boosted NLNR estimator (see Theorem~\ref*{theo-refine} in Section~3.2).

\medskip
For the weaker signal set \(\bm{M}_W^*\), we focus on a \emph{screening} guarantee.

\medskip
\noindent\textbf{Condition:}
\begin{enumerate}

\item[(Su2)]
There exists a constant $\widetilde{C}_{\mathrm{weak},H}>0$ such that
\[
\min\bigl\{|\beta_j^*|: j\in \bm{M}_W^{*}\bigr\}
\ \succ\
\tilde r_n,
\]
where $\tilde r_n=n^{-\gamma-\tilde{\alpha}}(\log \bar{s}\,\log n)^{1/2}$, and, uniformly over $j\in\bm{M}_W^{*}$,
\[
\left\|
\Bigl(\boldsymbol{X}_{\mathcal I_2,\widetilde H_j}^{\mathsf T}\boldsymbol{X}_{\mathcal I_2,\widetilde H_j}/|\mathcal I_2|\Bigr)^{-1}
\right\|_\infty
\le
c_{0,H}\,n^{1/2-\gamma-\tilde{\alpha}}(\log n)^{1/2}.
\]

\item[(Su3)] Let $d_{\max,W}=\max_{j\in \bm M_W^*} |\Omega_j|$. Assume
\[
|\bm M_W^*|\, d_{\max,W} = o(\bar s^{4}).
\]

\end{enumerate}

The following theorem establishes the sure screening property for $\bm M_W^*$ under the expanded working models.

\begin{theorem}\label{theo-pscre2-H}
  Suppose Assumption~\ref{assump:node} hold for $j\in\bm M_W^*$, and Conditions \textup{(Su2)}--\textup{(Su3)} hold.  Let $\tau_{nj}=\widetilde{C}_{\tau,j}\tilde r_n$ with $\widetilde{C}_{\tau,j}>0$ a constant. Then, for all sufficiently large $n$,
\[
\operatorname{pr}\!\left(
\bm{M}_W^{*}\subseteq \widetilde{\bm{M}}_{\boldsymbol{\tau}\mid\widetilde{H},\mathcal I_2}
\right)
\ge
1
-2\sum_{j\in \bm{M}_W^{*}} (|\Omega_j|+1)\,\bar{s}^{-4}
-o(p_n^{-2}).
\]
\end{theorem}

\medskip

\subsection{Supplementary Simulation Results}
We now examine model~(1) under varying levels of sparsity in the precision matrix $\boldsymbol{\Theta}$.
Specifically, we fix $p_n = 500$ and vary the correlation parameter $\rho \in \{0.1, 0.2, \ldots, 0.7\}$,
which directly controls the sparsity of $\boldsymbol{\Theta}$.
All other settings follow Section~5.1. Figure~S1 shows that NLNR and boosted NLNR remain robust across the entire range, with their advantage widening as \(\rho\) increases.

\setcounter{figure}{0}
\renewcommand{\thefigure}{S\arabic{figure}}

  \begin{figure}[hbpt]
    \centering
    \includegraphics[width=\textwidth]{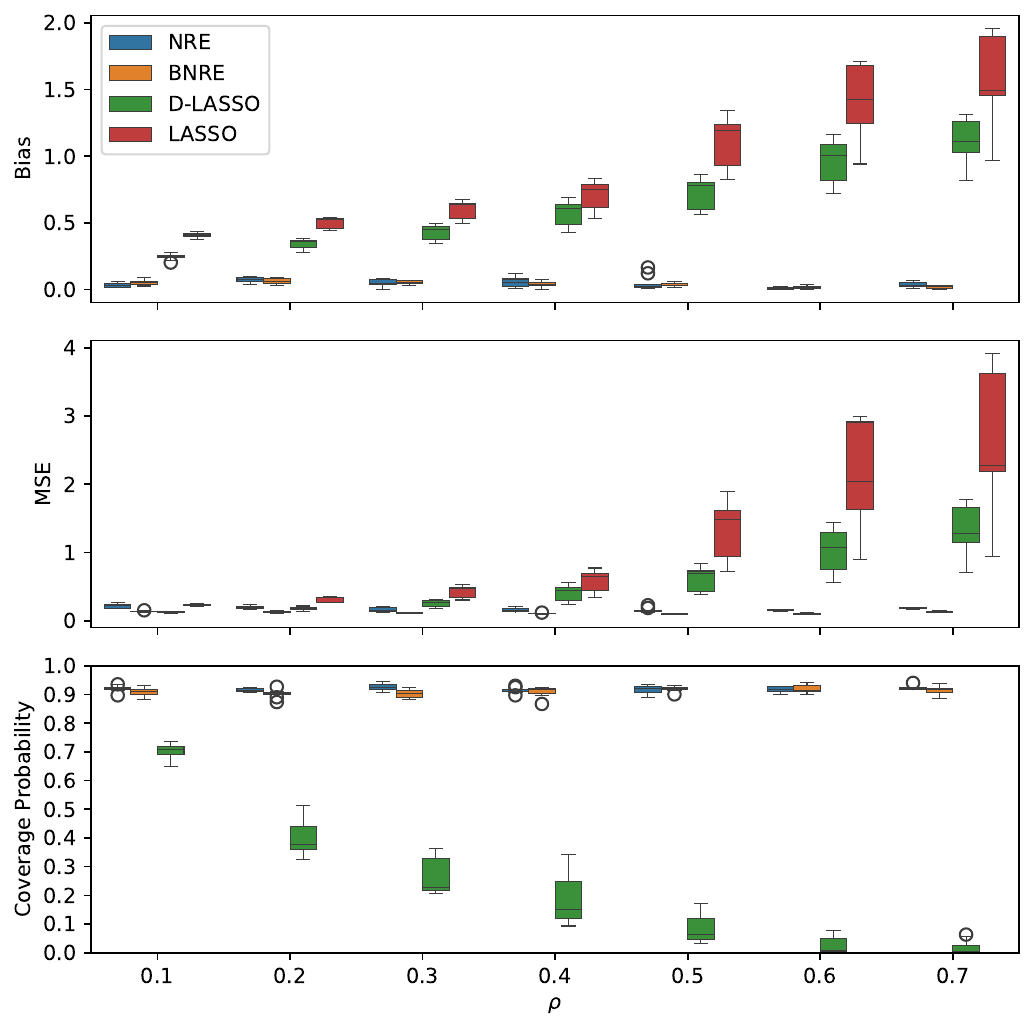}
    \caption{ Simulation results for the model~(1) with $p_n = 500$, $\beta_j = \beta_{20+j} = 3.5 - 0.5 j$ for $j = 1, 2, \dots, 5$, and $\beta_j = 0.2 \sin(j)$ for $j \notin \boldsymbol{M}^{*}=\{1{:}5,21{:}25\}$.
    The box plots display the distribution of each indicator across the strong signals $\{\beta_j: j \in \bm{M}^{*}\}$.
    The x-axis $\rho$ represents the sparsity level of $\bm{\Theta}$ as defined in Section 5.1.
    The sample sizes are $n = 200$ and $N - n = 6000$.
    }
  \end{figure}

Table~S1 provides additional variable selection results under the sparse and dense $\bm{\beta}$ configurations introduced in Section 5.1.

\setcounter{table}{0}
\renewcommand{\thetable}{S\arabic{table}}

  \begin{table}[hbpt]
    \centering
     \caption{Variable selection results under both sparse and dense $\boldsymbol{\beta}$ settings, as described in Section 5.1. The value of $\alpha_{\tau}$ denotes the threshold for the p-value in the variable selection procedure.
    The sample sizes are $n = 200$ and $N-n = 6000$.
    }
    \label{tab_varsel121}
    \resizebox{\textwidth}{!}{
      \begin{tabular}{llllllllll}
        \toprule
        \multirow{2}{*}{Method} &
        \multirow{2}{*}{$p_n$} &
        \multicolumn{4}{c}{Sparse $\boldsymbol{\beta}$} &
        \multicolumn{4}{c}{Dense $\boldsymbol{\beta}$} \\
        \cmidrule(lr){3-6} \cmidrule(lr){7-10}
        & & \multicolumn{2}{c}{$\alpha_{\tau}=10^{-3}$} & \multicolumn{2}{c}{$\alpha_{\tau}=10^{-4}$} & \multicolumn{2}{c}{$\alpha_{\tau}=10^{-3}$} & \multicolumn{2}{c}{$\alpha_{\tau}=10^{-4}$} \\
        \cmidrule(lr){3-4} \cmidrule(lr){5-6} \cmidrule(lr){7-8} \cmidrule(lr){9-10}
        & & FSR & NSR & FSR & NSR & FSR & NSR & FSR & NSR \\
        \midrule
        boosted NLNR       & 400  & 1.15E-2 & 0.00 & 3.16E-3 & 0.00 & 3.66E-2 & 0.00 & 1.36E-2 & 0.00 \\
        Debiased LASSO & 400  & 1.04E-2 & 3.17E-3 & 4.67E-3 & 4.50E-3 & 1.13E-1 & 2.33E-3 & 5.96E-2 & 3.00E-3 \\
        \addlinespace
        boosted NLNR       & 600  & 1.53E-2 & 0.00 & 5.63E-3 & 0.00 & 3.79E-2 & 6.67E-4 & 1.24E-2 & 1.67E-3 \\
        Debiased LASSO & 600  & 9.48E-3 & 2.43E-2 & 3.95E-3 & 3.37E-2 & 1.29E-2 & 5.02E-1 & 7.29E-3 & 5.69E-1 \\
        \addlinespace
        boosted NLNR       & 800  & 1.54E-2 & 0.00 & 5.96E-3 & 0.00 & 3.68E-2 & 4.33E-3 & 1.35E-2 & 1.08E-2 \\
        Debiased LASSO & 800  & 1.60E-2 & 8.73E-2 & 8.33E-3 & 1.07E-1 & 1.23E-2 & 6.52E-1 & 7.47E-3 & 7.12E-1 \\
        \addlinespace
        boosted NLNR       & 1000 & 2.06E-2 & 0.00 & 6.62E-3 & 0.00 & 3.98E-2 & 1.57E-2 & 1.66E-2 & 3.12E-2 \\
        Debiased LASSO & 1000 & 1.70E-2 & 1.80E-1 & 1.02E-2 & 2.09E-1 & 1.13E-2 & 7.97E-1 & 6.51E-3 & 8.47E-1 \\
        \addlinespace
        boosted NLNR       & 1200 & 2.34E-2 & 0.00 & 7.94E-3 & 0.00 & 3.90E-2 & 3.90E-2 & 1.63E-2 & 6.35E-2 \\
        Debiased LASSO & 1200 & 1.56E-2 & 2.54E-1 & 9.13E-3 & 2.77E-1 & 2.44E-2 & 7.67E-1 & 1.31E-2 & 8.12E-1 \\
        \addlinespace
        boosted NLNR       & 1400 & 2.09E-2 & 0.00 & 7.28E-3 & 0.00 & 4.57E-2 & 7.08E-2 & 1.50E-2 & 1.05E-1 \\
        Debiased LASSO & 1400 & 1.57E-2 & 3.10E-1 & 6.18E-3 & 3.30E-1 & 1.26E-2 & 8.70E-1 & 8.99E-3 & 9.08E-1 \\
        \addlinespace
        boosted NLNR       & 1600 & 2.25E-2 & 0.00 & 8.59E-3 & 0.00 & 4.41E-2 & 1.16E-1 & 1.68E-2 & 1.59E-1 \\
        Debiased LASSO & 1600 & 2.19E-2 & 3.22E-1 & 1.48E-2 & 3.45E-1 & 1.18E-2 & 8.89E-1 & 6.28E-3 & 9.21E-1 \\
        \bottomrule
    \end{tabular}
    }
  \end{table}

Tables~S2–S24 present the top 10 sensitivity genes identified across drugs, with the exception of Paclitaxel.


\begin{table}[htbp]
	\caption{Top 10 sensitive genes selected by boosted NLNR and D-LASSO for the drug 17-AAG}
	\label{tab_drug17AAG}
	\setlength{\tabcolsep}{3pt}
	\begin{tabular}{lccccc}
		\hline
		Rank & 1 & 2 & 3 & 4 & 5 \\ \hline
		boosted NLNR & NAPG & ZNF726 & RP11-138I17.1 & MB21D1 & HCN4 \\
		D-LASSO & NAPG & MB21D1 & PRKG2\_MUT & IFNA5 & PGM2L1 \\
		\hline
		Rank & 6 & 7 & 8 & 9 & 10 \\ \hline
		boosted NLNR & ZNF585A & PGM2L1 & CTB-92J24.3 & GAL3ST3 & BAX \\
		D-LASSO & RP11-339F13.2 & ZNF726 & CTDSP1 & ARHGAP29 & RP11-468N14.09 \\ \hline
	\end{tabular}	
\end{table}

\begin{table}[htbp]
	\caption{Top 10 sensitive genes selected by boosted NLNR and D-LASSO for the drug AEW541}
	\label{tab_drugAEW541}
	\begin{tabular}{lccccc}
		\hline
		Rank & 1 & 2 & 3 & 4 & 5 \\ \hline
		boosted NLNR & IGLV3-6 & IFNG-AS1 & GLRXP3 & MYOM2 & DERL3 \\
		D-LASSO & KCNAB1 & SLC44A1 & NUP210P3 & MYOM2 & TNRC18P2 \\
		\hline
		Rank & 6 & 7 & 8 & 9 & 10 \\ \hline
		boosted NLNR & IGLCOR22-1 & TNRC18P2 & IGHA1 & NUP210P3 & RP11-325F22.2 \\
		D-LASSO & AL928742.1 & TMEM101 & COMMD3 & ATP11B & VPS51 \\ \hline
	\end{tabular}
\end{table}

\begin{table}[htbp]
	\caption{Top 10 sensitive genes selected by boosted NLNR and D-LASSO for the drug AZD0530}
	\label{tab_drugAZD0530}
	\setlength{\tabcolsep}{2pt}
	\begin{tabular}{lccccc}
		\hline
		Rank & 1 & 2 & 3 & 4 & 5 \\ \hline
		boosted NLNR & TRGC2 & TRGJ2 & TRGJ1 & PIM1 & CDRT15P2 \\
		D-LASSO & CTSV & PRTN3 & CEBPE & PIM1 & BTC \\
	\hline
		Rank & 6 & 7 & 8 & 9 & 10 \\ \hline
		boosted NLNR & GGTLC1 & XXbac-BPG13B8.10 & JRKL-AS1 & CTSV & PRTN3 \\
		D-LASSO & XXbac-BPG13B8.10 & ZNF581 & TMEM229B & MUC22 & PCNP \\ \hline
	\end{tabular}
	
\end{table}


\begin{table}[htbp]
	\caption{Top 10 sensitive genes selected by boosted NLNR and D-LASSO for the drug AZD6244}
	\label{tab_drugAZD6244}
	\setlength{\tabcolsep}{3pt}
	\begin{tabular}{lccccc}
			\hline
		Rank & 1 & 2 & 3 & 4 & 5 \\ 	\hline
		boosted NLNR & RNASE2 & RXRG & FCGR1A & PPP1R27 & DPRXP2 \\
		D-LASSO & NRAS\_MUT & RP11-291H24.1 & ZFP36L2 & BRAF.V600E\_MUT & SPRY2 \\
			\hline
		Rank & 6 & 7 & 8 & 9 & 10 \\ 	\hline
		boosted NLNR & RP11-403I13.9 & KRAS & C19orf59 & RP11-380F14.2 & ZFP36L2 \\
		D-LASSO & RP11-366L5.1 & QDPR & MGST2 & RP11-57A1.1 & PYCARD \\ 	\hline
	\end{tabular}
	
\end{table}

\begin{table}[htbp]
	\caption{Top 10 sensitive genes selected by boosted NLNR and D-LASSO for the drug Erlotinib}
	\label{tab_drugErlotinib}
	\setlength{\tabcolsep}{3pt}
	\small
	\begin{tabular}{lccccc}
		\hline
		Rank & 1 & 2 & 3 & 4 & 5 \\ \hline
		boosted NLNR & CTD-2117L12.1 & MUC22 & GPR116 & CTSV & GGTLC1 \\
		D-LASSO & SOX15 & CTSV & RP11-338E21.2 & CLK3\_MUT & CNKSR1 \\
		\hline
		Rank & 6 & 7 & 8 & 9 & 10 \\ \hline
		boosted NLNR & ELOVL7 & RP11-195E2.1 & CLK3\_MUT & RP11-902B17.1 & RP11-770G2.4 \\
		D-LASSO & XKRX & TPRXL & SLCO2A1 & MYO16-AS1 & SH3YL1 \\ \hline
	\end{tabular}
	
\end{table}

\begin{table}[htbp]
	\caption{Top 10 sensitive genes selected by boosted NLNR and D-LASSO for the drug Irinotecan}
	\label{tab_drugIrinotecan}
	\begin{tabular}{lccccc}
		\hline
		Rank & 1 & 2 & 3 & 4 & 5 \\ \hline
		boosted NLNR & SLFN11 & CTSD & STX18-AS1 & PSTPIP1 & LSM11 \\
		D-LASSO & SLFN11 & VDAC1P9 & CETN2 & SART3 & SLC35F6 \\
		\hline
		Rank & 6 & 7 & 8 & 9 & 10 \\ \hline
		boosted NLNR & LMNB1 & CSK & HNRNPA1P39 & ARHGAP19 & MYCBP2-AS1 \\
		D-LASSO & KHDRBS1 & PPP1R8 & LMNB1 & ARHGAP19 & RP11-404F10.2 \\ \hline
	\end{tabular}
	
\end{table}

\begin{table}[htbp]
	\caption{Top 10 sensitive genes selected by boosted NLNR and D-LASSO for the drug L-685458}
	\label{tab_drugL685458}
	\setlength{\tabcolsep}{4.2pt}
	\begin{tabular}{lccccc}
			\hline
		Rank & 1 & 2 & 3 & 4 & 5 \\ 	\hline
		boosted NLNR & ASZ1 & VDAC1P9 & P2RX1 & RP11-480N24.3 & NLRC3 \\
		D-LASSO & TRGV5P & CDRT15L2 & RP11-480N24.3 & RNU4-34P & RETN \\
			\hline
		Rank & 6 & 7 & 8 & 9 & 10 \\ 	\hline
		boosted NLNR & IKZF1 & LINC00528 & RP11-455F5.5 & FAM129B & CDT1 \\
		D-LASSO & RP11-12A2.1 & NLRC3 & RP1-225E12.2 & WAS & ARHGAP30 \\ 	\hline
	\end{tabular}
	
\end{table}

\begin{table}[htbp]
	\caption{Top 10 sensitive genes selected by boosted NLNR and D-LASSO for the drug LBW242}
	\label{tab_drugLBW242}
		\setlength{\tabcolsep}{2pt} %
	\begin{tabular}{lccccc}
			\hline
		Rank & 1 & 2 & 3 & 4 & 5 \\ 	\hline
		boosted NLNR & PDPK1\_MUT & EFCAB3 & CTD-2378E21.1 & RP11-149B9.2 & RP11-562F9.2 \\
		D-LASSO & CXCL1 & CTBP2P2 & TULP3P1 & METTL7AP1 & AC005162.5 \\
			\hline
		Rank & 6 & 7 & 8 & 9 & 10 \\ 	\hline
		boosted NLNR & RP11-168O16.2 & CTC-756D1.1 & MRC1 & IL17F & RP11-234K19.1 \\
		D-LASSO & NCAM1\_MUT & OR5AP1P & RIPK1 & RN7SL442P & CPSF4L \\ 	\hline
	\end{tabular}
	
\end{table}

\begin{table}[htbp]
	\caption{Top 10 sensitive genes selected by boosted NLNR and D-LASSO for the drug Lapatinib}
	\label{tab_drugLapatinib}
		\setlength{\tabcolsep}{3pt} %
	\begin{tabular}{lccccc}
		\hline
		Rank & 1 & 2 & 3 & 4 & 5 \\ \hline
		boosted NLNR & LSR & ADIRF & RP11-4K16.2 & KB-1183D5.16 & PRSS8 \\
		D-LASSO & CTB-191K22.5 & VN1R21P & MYH14 & PRSS8 & PGAP3 \\
		\hline
		Rank & 6 & 7 & 8 & 9 & 10 \\ \hline
		boosted NLNR & RBBP8NL & RP11-195E2.1 & SMIM22 & RP11-354M1.2 & RP11-388M20.2 \\
		D-LASSO & KB-1183D5.16 & CDH1 & Z83851.4 & TMPRSS13 & MAL2 \\ \hline
	\end{tabular}

\end{table}

\begin{table}[htbp]
	\caption{Top 10 sensitive genes selected by boosted NLNR and D-LASSO for the drug Nilotinib.}
	\label{tab_drugNilotinib}
		\setlength{\tabcolsep}{3pt} %
		\small
	\begin{tabular}{lccccc}
		\hline
		Rank & 1 & 2 & 3 & 4 & 5 \\ \hline
		boosted NLNR & SNORA26 & RP11-234K19.1 & XXbac-BPG13B8.10 & SOD1P3 & LINC00861 \\
		D-LASSO & IL2RG & SIGLEC5 & ITGA4 & CTC-546K23.1 & RPL9P33 \\
	\hline
		Rank & 6 & 7 & 8 & 9 & 10 \\ \hline
		boosted NLNR & KLF1 & CRISP3 & FGF23 & AC092484.1 & CTC-546K23.1 \\
		D-LASSO & TTTY13 & OR14L1P & PTPN7 & MAP2K7 & SLC26A8 \\\hline
	\end{tabular}
	
\end{table}

\begin{table}[htbp]
	\caption{Top 10 sensitive genes selected by boosted NLNR and D-LASSO for the drug Nutlin-3.}
	\label{tab_drugNutlin3}
		\setlength{\tabcolsep}{3pt} %
	\begin{tabular}{lccccc}
			\hline
		Rank & 1 & 2 & 3 & 4 & 5 \\ 	\hline
		boosted NLNR & AC068286.1 & MS4A4A & AC005324.7 & RP11-255H23.4 & AC011897.2 \\
		D-LASSO & RP11-255H23.4 & AC005324.7 & RP11-149P24.1 & IGHM & LILRB5 \\
			\hline
		Rank & 6 & 7 & 8 & 9 & 10 \\ 	\hline
		boosted NLNR & SELL & C1QC & AC007680.2 & C16orf54 & AC011515.2 \\
		D-LASSO & AIP & TP53\_MUT & AC023115.4 & FAM78A & OR2T35 \\ 	\hline
	\end{tabular}
\end{table}

\begin{table}[htbp]
	\caption{Top 10 sensitive genes selected by boosted NLNR and D-LASSO for the drug PD-0325901.}
	\label{tab_drugPD0325901}
	\centering
		\setlength{\tabcolsep}{8pt} %
	\begin{tabular}{lccccc}
		\hline
		Rank & 1 & 2 & 3 & 4 & 5 \\ \hline
		boosted NLNR & RXRG & HSD17B11 & QDPR & MGST2 & RP4-705D16.3 \\
		D-LASSO & SPRY2 & MGST2 & CD63 & SGK1 & RP11-53I6.2 \\
		\hline
		Rank & 6 & 7 & 8 & 9 & 10 \\ \hline
		boosted NLNR & C10orf90 & AP1S2 & RP11-529H2.2 & PLP1 & RP11-53I6.2 \\
		D-LASSO & ZCCHC9 & DPP7 & QDPR & SDC3 & MIR4497 \\\hline 
	\end{tabular} 
	
\end{table}

\begin{table}[htbp]
	\caption{Top 10 sensitive genes selected by boosted NLNR and D-LASSO for the drug PD-0332991.}
	\label{tab_drugPD0332991}
	\centering
		\setlength{\tabcolsep}{4.5pt} %
	\begin{tabular}{lccccc}
		\hline
		Rank & 1 & 2 & 3 & 4 & 5 \\ \hline
		boosted NLNR & LIPJ & RB1\_MUT & AMER2-AS1 & PTEN & CECR1 \\
		D-LASSO & LIPJ & CDRT15L2 & AMER2-AS1 & AC006994.1 & AL162151.3 \\
	\hline
		Rank & 6 & 7 & 8 & 9 & 10 \\ \hline
		boosted NLNR & LINC00957 & CDRT15L2 & TNFAIP8 & AC006994.1 & AGAP2 \\
		D-LASSO & MEF2C & RB1\_MUT & CECR1 & RP11-264I13.2 & RP11-408P14.1 \\ \hline
	\end{tabular}
	\label{tab_drugPD0332991}
	
\end{table}

\begin{table}[htbp]
	\caption{Top 10 sensitive genes selected by boosted NLNR and D-LASSO for the drug PF2341066.}
	\label{tab_drugPF2341066}
		\setlength{\tabcolsep}{3pt} %
	\centering
	\begin{tabular}{lccccc}
		\hline
		Rank & 1 & 2 & 3 & 4 & 5 \\ \hline
		boosted NLNR & SELPLG & RP11-453O22.1 & RP11-168O16.1 & RP11-659P15.2 & WAS \\
		D-LASSO & RP11-61D1.2 & WAS & HGF & NLRC4 & CEBPE \\
	\hline
		Rank & 6 & 7 & 8 & 9 & 10 \\ \hline
		boosted NLNR & CLDN24 & GPSM3 & NLRC4 & CEBPE & ITGAL \\
		D-LASSO & MCM7 & SELPLG & RP11-550P17.5 & MET & PTGDR2 \\ \hline
	\end{tabular}
	
\end{table}

\begin{table}[htbp]
	\caption{Top 10 sensitive genes selected by boosted NLNR and D-LASSO for the drug PHA-665752}
	\label{tab_drugPHA665752}
		\setlength{\tabcolsep}{3pt} %
	\begin{tabular}{lccccc}
		\hline
		Rank & 1 & 2 & 3 & 4 & 5 \\ \hline
		boosted NLNR & TRBV27 & RP11-562F9.2 & CTC-295J13.3 & ZNF683 & LINC00989 \\
		D-LASSO & OR5BA1P & RP11-123H22.1 & GRK4\_MUT & LINC00400 & RP5-1051H14.2 \\
		\hline
		Rank & 6 & 7 & 8 & 9 & 10 \\ \hline
		boosted NLNR & TRGJP2 & TRGC2 & RP11-234K19.1 & PTPRC & RN7SL646P \\
		D-LASSO & AMER2-AS1 & CTDP1 & MICB & UBR2 & VDAC1P9 \\ \hline
	\end{tabular}
	
\end{table}

\begin{table}[htbp]
	\centering
	\caption{Top 10 sensitive genes selected by boosted NLNR and D-LASSO for the drug PLX4720}
	\label{tab_drugPLX4720}
	\setlength{\tabcolsep}{1.8pt} %
	\small
	\begin{tabular}{lccccc}
		\hline
		Rank & 1 & 2 & 3 & 4 & 5 \\ \hline
		boosted NLNR & LAMA1 & RXRG & RP11-76K13.3 & PLP1 & KRAS \\
		D-LASSO & BRAF.V600E\_MUT & GAPDHS & PTMAP6 & IRAK1\_MUT & RP11-726G1.1 \\
		\hline
		Rank & 6 & 7 & 8 & 9 & 10 \\ \hline
		boosted NLNR & HCCAT5 & CDRT15L2 & RP4-718J7.4 & RNASE2 & BRAF.V600E\_MUT \\
		D-LASSO & HCCAT5 & RP5-1195D24.1 & RP4-569M23.2 & GNPTAB & SALL4P1 \\ \hline
	\end{tabular}
\end{table}

\begin{table}[htbp]
	\caption{Top 10 sensitive genes selected by boosted NLNR and D-LASSO for the drug Panobinostat}
	\label{tab_drugPanobinostat}
	\begin{tabular}{lccccc}
		\hline
		Rank & 1 & 2 & 3 & 4 & 5 \\ \hline
		boosted NLNR & TAF3 & EAF2 & MYCBP2-AS1 & AC004490.1 & EIF4EBP2 \\
		D-LASSO & MYCBP2-AS1 & EIF4EBP2 & TAF3 & TNFAIP8 & SFMBT2 \\
		\hline
		Rank & 6 & 7 & 8 & 9 & 10 \\ \hline
		boosted NLNR & MIR142 & DENND4B & RAD51 & RASA4CP & PUM2 \\
		D-LASSO & BLM & FTH1P7 & NCKAP1 & RASA4CP & MYCBP2 \\\hline
	\end{tabular}
	
\end{table}

\begin{table}[htbp]
	\caption{Top 10 sensitive genes selected by boosted NLNR and D-LASSO for the drug RAF265}
	\label{tab_drugRAF265}
	\begin{tabular}{lccccc}
			\hline
		Rank & 1 & 2 & 3 & 4 & 5 \\ 	\hline
		boosted NLNR & AVP & RP11-163O19.1 & MNDA & RNASE2 & SERPINB10 \\
		D-LASSO & GNPTAB & CTD-2574D22.4 & FAM71D & EIF2AK4 & CMTM3 \\
			\hline
		Rank & 6 & 7 & 8 & 9 & 10 \\ 	\hline
		boosted NLNR & TRBV27 & RP11-84C10.2 & CLEC4C & ZNF683 & DUX4L18 \\
		D-LASSO & CDRT15L2 & RN7SL64P & GIT2 & PARP8 & TFAP2E \\ 	\hline
	\end{tabular}
	
\end{table}

\begin{table}[htbp]
	\caption{Top 10 sensitive genes selected by boosted NLNR and D-LASSO for the drug Sorafenib}
	\label{tab_drugSorafenib}
		\setlength{\tabcolsep}{3pt} %
	\begin{tabular}{lccccc}
			\hline
		Rank & 1 & 2 & 3 & 4 & 5 \\ 	\hline
		boosted NLNR & RNASE2 & KRAS & LST1 & RP11-26F2.2 & RP11-480N24.3 \\
		D-LASSO & RP11-480N24.3 & RN7SL646P & EMR4P & PROK2 & LPPR3 \\
			\hline
		Rank & 6 & 7 & 8 & 9 & 10 \\ 	\hline
		boosted NLNR & SIGLEC5 & MNDA & CTB-138E5.1 & NKG7 & RP4-728D4.2 \\
		D-LASSO & EIF3L & C9orf47 & RP11-12A2.1 & NMUR1 & LINC00528 \\ 		\hline
	\end{tabular}

\end{table}

\begin{table}[htbp]
	\caption{Top 10 sensitive genes selected by boosted NLNR and D-LASSO for the drug TAE684}
	\label{tab_drugTAE684}
		\setlength{\tabcolsep}{3pt} %
	\begin{tabular}{lccccc}
		\hline
		Rank & 1 & 2 & 3 & 4 & 5 \\ \hline
		boosted NLNR & OCSTAMP & RP11-153M7.1 & RP11-734C14.2 & WNT9B & TMPRSS11F \\
		D-LASSO & AC003005.2 & ARHGAP30 & PIM2 & AMER2-AS1 & AC139099.6 \\
		\hline
		Rank & 6 & 7 & 8 & 9 & 10 \\ \hline
		boosted NLNR & FTLP10 & RP11-453O22.1 & SUPT20HL2 & SELPLG & ENPP6 \\
		D-LASSO & ARID3A & ENPP6 & CSNK1G2 & STON1 & C2orf43 \\ \hline
	\end{tabular}
\end{table}

\begin{table}[htbp]
	\caption{Top 10 sensitive genes selected by boosted NLNR and D-LASSO for the drug TKI258}
	\label{tab_drugTKI258}
	\begin{tabular}{lccccc}
		\hline
		Rank & 1 & 2 & 3 & 4 & 5 \\ \hline
		boosted NLNR & LYL1 & RP11-26F2.2 & RP11-1180F24.1 & MS4A14 & AC005324.7 \\
		D-LASSO & SNX20 & TYMS & RILPL2 & IQGAP2 & PDE7A \\
		\hline
		Rank & 6 & 7 & 8 & 9 & 10 \\ \hline
		boosted NLNR & FCGR1C & RILPL2 & RP11-544A12.5 & PROK2 & RNASE2 \\
		D-LASSO & PROSC & MYB & DPEP2 & C9orf47 & FRY \\\hline
	\end{tabular}
	
\end{table}

\begin{table}[htbp]
	\centering
	\caption{Top 10 sensitive genes selected by boosted NLNR and D-LASSO for the drug Topotecan}
	\label{tab_drugTopotecan}
		\setlength{\tabcolsep}{13pt} %
	\begin{tabular}{lccccc}
		 \hline
		Rank & 1 & 2 & 3 & 4 & 5 \\  \hline
		boosted NLNR & SLFN11 & MAGED2 & AGAP2 & SF3A2 & LMNB1 \\
		D-LASSO & SLFN11 & MAGED2 & LMNB1 & MCM3 & CDC40 \\
		 \hline
		Rank & 6 & 7 & 8 & 9 & 10 \\
		boosted NLNR & DOCK2 & SLC35A2 & CDC40 & C4orf46 & ABCB1 \\
		D-LASSO & C4orf46 & ABCB1 & SLC35A2 & TMA16 & SF3A2 \\  \hline
	\end{tabular}
	
\end{table}

\begin{table}[htbp]
	\caption{Top 10 sensitive genes selected by boosted NLNR and D-LASSO for the drug ZD-6474}
	\label{tab_drugZD6474}
		\setlength{\tabcolsep}{4pt} %
	\small
	\begin{tabular}{lccccc}
		\hline
		Rank & 1 & 2 & 3 & 4 & 5 \\ 	\hline
		boosted NLNR & GGTLC1 & METTL15P3 & AMY1B & METTL11B & RP11-902B17.1 \\
		D-LASSO & REL\_MUT & CTSV & RP11-301G19.1 & SYDE1 & RP1-68D18.3 \\
	 \hline
		Rank & 6 & 7 & 8 & 9 & 10 \\ 	\hline
		boosted NLNR & RP11-301G19.1 & METTL21C & CXCL11 & RPS2P52 & CTSV \\
		D-LASSO & CXCL11 & RP11-96C23.10 & IQGAP2 & COL13A1 & EGFR\_MUT \\
		\hline
	\end{tabular}
	
\end{table}

\vskip 0.2in
\bibliography{NR-ref1}

@article{Alalawy2024Key,
  author  = {Alalawy, A. I.},
  title   = {Key genes and molecular mechanisms related to {Paclitaxel} resistance},
  journal = {Cancer Cell International},
  year    = {2024},
  volume  = {24},
  number  = {1},
  pages   = {244}
}

@article{Bao2024,
  author  = {Bao, Y. and Huo, Y. and Ren, H. and Zou, C.},
  title   = {Selective conformal inference with false coverage-statement rate control},
  journal = {Biometrika},
  year    = {2024},
  volume  = {111},
  number  = {3},
  pages   = {727--742}
}

@article{Barretina2012,
  author  = {Barretina, J. and Caponigro, G. and Stransky, N. and Venkatesan, K. and Margolin, A. A. and Kim, S. and others},
  title   = {The cancer cell line encyclopedia enables predictive modelling of anticancer drug sensitivity},
  journal = {Nature},
  year    = {2012},
  volume  = {483},
  number  = {7391},
  pages   = {603--607}
}

@article{Bickel2009,
  author  = {Bickel, P. J. and Ritov, Y. and Tsybakov, A. B.},
  title   = {Simultaneous analysis of {LASSO} and {Dantzig} selector},
  journal = {The Annals of Statistics},
  year    = {2009},
  volume  = {37},
  number  = {4},
  pages   = {1705--1732}
}

@article{Boyle2017,
  author  = {Boyle, E. A. and Li, Y. I. and Pritchard, J. K.},
  title   = {An expanded view of complex traits: from polygenic to omnigenic},
  journal = {Cell},
  year    = {2017},
  volume  = {169},
  number  = {7},
  pages   = {1177--1186}
}

@article{Bradic2022,
  author  = {Bradic, J. and Fan, J. and Zhu, Y.},
  title   = {Testability of high-dimensional linear models with nonsparse structures},
  journal = {The Annals of Statistics},
  year    = {2022},
  volume  = {50},
  number  = {2},
  pages   = {615--640}
}

@article{cai2011constrained,
  author  = {Cai, T. and Liu, W. and Luo, X.},
  title   = {A constrained $\ell_1$-minimization for sparse precision matrix estimation},
  journal = {Journal of the American Statistical Association},
  year    = {2011},
  volume  = {106},
  number  = {494},
  pages   = {594--607}
}

@article{Condorelli2018Polyclonal,
  author  = {Condorelli, R. and Spring, L. and O'Shaughnessy, J. and Lacroix, L. and Bailleux, C. and Scott, V. and Dubois, J. and Nagy, R. J. and Lanman, R. B. and Iafrate, A. J. and Andr{\'e}, F. and Bardia, A.},
  title   = {Polyclonal {RB1} mutations and acquired resistance to {CDK} 4/6 inhibitors in patients with metastatic breast cancer},
  journal = {Annals of Oncology},
  year    = {2018},
  volume  = {29},
  number  = {3},
  pages   = {640--645}
}

@article{Cook2012,
  author  = {Cook, R. D. and Forzani, L. and Rothman, A. J.},
  title   = {Estimating sufficient reductions of the predictors in abundant high-dimensional regressions},
  journal = {The Annals of Statistics},
  year    = {2012},
  volume  = {40},
  number  = {1},
  pages   = {353--384}
}

@article{du2023,
  author  = {Du, J. H. and Guo, Y. and Wang, X.},
  title   = {High-dimensional portfolio selection with cardinality constraints},
  journal = {Journal of the American Statistical Association},
  year    = {2023},
  volume  = {118},
  number  = {542},
  pages   = {779--791}
}

@article{Fan2001,
  author  = {Fan, J. and Li, R.},
  title   = {Variable selection via nonconcave penalized likelihood and its oracle properties},
  journal = {Journal of the American Statistical Association},
  year    = {2001},
  volume  = {96},
  number  = {456},
  pages   = {1348--1360}
}

@article{FanLv2008,
  author  = {Fan, J. and Lv, J.},
  title   = {Sure independence screening for ultrahigh-dimensional feature space},
  journal = {Journal of the Royal Statistical Society, Series B (Statistical Methodology)},
  year    = {2008},
  volume  = {70},
  number  = {5},
  pages   = {849--911}
}

@article{FanLv2010,
  author  = {Fan, J. and Lv, J.},
  title   = {A selective overview of variable selection in high-dimensional feature space},
  journal = {Statistica Sinica},
  year    = {2010},
  volume  = {20},
  number  = {1},
  pages   = {101--148}
}

@article{FanSong2010,
  author  = {Fan, J. and Song, R.},
  title   = {Sure independence screening in generalized linear models with NP-dimensionality},
  journal = {The Annals of Statistics},
  year    = {2010},
  volume  = {38},
  number  = {6},
  pages   = {3567--3604}
}

@article{Fan2011,
  author  = {Fan, J. and Lv, J.},
  title   = {Nonconcave penalized likelihood with NP-dimensionality},
  journal = {IEEE Transactions on Information Theory},
  year    = {2011},
  volume  = {57},
  number  = {8},
  pages   = {5467--5484}
}

@article{fan2014challenges,
  author  = {Fan, J. and Han, F. and Liu, H.},
  title   = {Challenges of big data analysis},
  journal = {National Science Review},
  year    = {2014},
  volume  = {1},
  number  = {2},
  pages   = {293--314}
}

@article{Fan2016,
  author  = {Fan, J. and Liao, Y. and Liu, H.},
  title   = {An overview of the estimation of large covariance and precision matrices},
  journal = {The Econometrics Journal},
  year    = {2016},
  volume  = {19},
  number  = {1},
  pages   = {C1--C32}
}

@article{Bo2014paclitaxel,
  author  = {Gao, B. and Russell, A. and Beesley, J. and Chen, X. and Healey, S. and Henderson, M. and others},
  title   = {{Paclitaxel} sensitivity in relation to {ABCB1} expression, efflux and single nucleotide polymorphisms in ovarian cancer},
  journal = {Scientific Reports},
  year    = {2014},
  volume  = {4},
  number  = {1},
  pages   = {4669}
}

@article{Gao2012TUBB3,
  author  = {Gao, S. and Zhao, X. and Lin, B. and Hu, Z. and Yan, L. and Gao, J.},
  title   = {Clinical implications of {REST} and {TUBB3} in ovarian cancer and its relationship to paclitaxel resistance},
  journal = {Tumor Biology},
  year    = {2012},
  volume  = {33},
  number  = {5},
  pages   = {1759--1765}
}

@article{Giannone2022,
  author  = {Giannone, D. and Lenza, M. and Primiceri, G. E.},
  title   = {Economic predictions with big data: the illusion of sparsity},
  journal = {Econometrica},
  year    = {2022},
  volume  = {89},
  number  = {5},
  pages   = {2409--2437}
}

@article{Holm1979,
  author  = {Holm, S.},
  title   = {A simple sequentially rejective multiple test procedure},
  journal = {Scandinavian Journal of Statistics},
  year    = {1979},
  volume  = {6},
  number  = {2},
  pages   = {65--70}
}

@article{Huang2008Bridge,
  author  = {Huang, J. and Horowitz, J. L. and Ma, S.},
  title   = {Asymptotic properties of bridge estimators in sparse high-dimensional regression models},
  journal = {The Annals of Statistics},
  year    = {2008},
  volume  = {36},
  number  = {2},
  pages   = {587--613}
}

@article{Huang2009GroupBridge,
  author  = {Huang, J. and Ma, S. and Xie, H. and Zhang, C. H.},
  title   = {A group bridge approach for variable selection},
  journal = {Biometrika},
  year    = {2009},
  volume  = {96},
  number  = {2},
  pages   = {339--355}
}

@article{Javanmard2014,
  author  = {Javanmard, A. and Montanari, A.},
  title   = {Confidence intervals and hypothesis testing for high-dimensional regression},
  journal = {Journal of Machine Learning Research},
  year    = {2014},
  volume  = {15},
  number  = {1},
  pages   = {2869--2909}
}

@article{Lee2016,
  author  = {Lee, J. D. and Sun, D. L. and Sun, Y. and Taylor, J. E.},
  title   = {Exact post-selection inference, with application to the {LASSO}},
  journal = {The Annals of Statistics},
  year    = {2016},
  volume  = {44},
  number  = {3},
  pages   = {907--927}
}

@article{LiG2012,
  author  = {Li, G. and Heng, P. and Zhang, J. and Zhu, L.},
  title   = {Robust rank correlation-based screening},
  journal = {The Annals of Statistics},
  year    = {2012},
  volume  = {40},
  number  = {3},
  pages   = {1846--1877}
}

@article{LiR2012,
  author  = {Li, R. and Zhong, W. and Zhu, L.},
  title   = {Feature screening via distance correlation learning},
  journal = {Journal of the American Statistical Association},
  year    = {2012},
  volume  = {107},
  number  = {499},
  pages   = {1129--1139}
}

@article{Liang2022,
  author  = {Liang, F. and Xue, J. and Jia, B.},
  title   = {Markov neighborhood regression for high-dimensional inference},
  journal = {Journal of the American Statistical Association},
  year    = {2022},
  volume  = {117},
  number  = {539},
  pages   = {1200--1214}
}

@article{Li2022Case,
  author  = {Li, J. and Wei, B. and Feng, J. and Wu, X. and Chang, Y. and Wang, Y. and others},
  title   = {Case report: {TP53} and {RB1} loss may facilitate the transformation from lung adenocarcinoma to small cell lung cancer by expressing neuroendocrine markers},
  journal = {Frontiers in Endocrinology},
  year    = {2022},
  volume  = {13},
  pages   = {1006480}
}

@article{Liu2019DrugResistance,
  author  = {Liu, H. and Wang, S. and Zhou, S. and Meng, Q. and Ma, X. and Song, X. and Wang, L. and Jiang, W.},
  title   = {Drug resistance-related competing interactions of lnc{RNA} and m{RNA} across 19 cancer types},
  journal = {Molecular Therapy: Nucleic Acids},
  year    = {2019},
  volume  = {16},
  pages   = {442--451}
}

@article{Meinshausen2006,
  author  = {Meinshausen, N. and B{\"u}hlmann, P.},
  title   = {High-dimensional graphs and variable selection with the {LASSO}},
  journal = {The Annals of Statistics},
  year    = {2006},
  volume  = {34},
  number  = {3},
  pages   = {1436--1462}
}

@article{Mozzetti2005,
  author  = {Mozzetti, S. and Ferlini, C. and Concolino, P. and Filippetti, F. and Raspaglio, G. and Prislei, S. and Gallo, D. and Martinelli, E. and Ranelletti, F. O. and Ferrandina, G. and Scambia, G.},
  title   = {Class {III} beta-tubulin overexpression is a prominent mechanism of paclitaxel resistance in ovarian cancer patients},
  journal = {Clinical Cancer Research},
  year    = {2005},
  volume  = {11},
  number  = {1},
  pages   = {298--305}
}

@article{Song2015,
  author  = {Song, Q. and Liang, F.},
  title   = {High-dimensional variable selection with reciprocal {L1}-regularization},
  journal = {Journal of the American Statistical Association},
  year    = {2015},
  volume  = {110},
  number  = {512},
  pages   = {1607--1620}
}

@article{Salmena2011ceRNA,
  author  = {Salmena, L. and Poliseno, L. and Tay, Y. and Kats, L. and Pandolfi, P.},
  title   = {A {ceRNA} hypothesis: the {Rosetta} {Stone} of a hidden {RNA} language?},
  journal = {Cell},
  year    = {2011},
  volume  = {146},
  number  = {3},
  pages   = {353--358}
}

@article{Sullivan2011BRAF,
  author  = {Sullivan, R. J. and Flaherty, K. T.},
  title   = {{BRAF} in melanoma: pathogenesis, diagnosis, inhibition, and resistance},
  journal = {Journal of Skin Cancer},
  year    = {2011},
  volume  = {2011},
  pages   = {423239}
}

@article{Sun2019CDC25A,
  author  = {Sun, Y. and Li, S. and Yang, L. and Zhang, D. and Zhao, Z. and Gao, J. and Liu, L.},
  title   = {{CDC25A} facilitates chemo-resistance in ovarian cancer multicellular spheroids by promoting e-cadherin expression and arresting cell cycles},
  journal = {Journal of Cancer},
  year    = {2019},
  volume  = {10},
  pages   = {2874--2884}
}

@article{Song2025Crotonylation,
  author  = {Song, H. and Guo, Z. and Xie, K. and Liu, X. and Yang, X. and Shen, R. and Wang, D.},
  title   = {Crotonylation of {MCM6} enhances chemotherapeutics sensitivity of breast cancer via inducing {DNA} replication stress},
  journal = {Cell Proliferation},
  year    = {2025},
  volume  = {58},
  number  = {2},
  pages   = {e13759}
}

@article{James2008Discovery,
  author  = {Tsai, J. and Lee, J. T. and Wang, W. and Zhang, J. and Cho, H. and Mamo, S. and others},
  title   = {Discovery of a selective inhibitor of oncogenic {BRAF} kinase with potent antimelanoma activity},
  journal = {Proceedings of the National Academy of Sciences of the United States of America},
  year    = {2008},
  volume  = {105},
  number  = {8},
  pages   = {3041--3046}
}

@article{Tibshirani1996,
  author  = {Tibshirani, R.},
  title   = {Regression shrinkage and selection via the {LASSO}},
  journal = {Journal of the Royal Statistical Society, Series B (Statistical Methodology)},
  year    = {1996},
  volume  = {58},
  number  = {1},
  pages   = {267--288}
}

@article{VandeGeer2014,
  author  = {Van de Geer, S. and B{\"u}hlmann, P. and Ritov, Y. and Dezeure, R.},
  title   = {On asymptotically optimal confidence regions and tests for high-dimensional models},
  journal = {The Annals of Statistics},
  year    = {2014},
  volume  = {42},
  number  = {3},
  pages   = {1166--1202}
}

@article{Wainwright2009,
  author  = {Wainwright, M. J.},
  title   = {Sharp thresholds for high-dimensional and noisy sparsity recovery using $\ell_{1}$-constrained quadratic programming ({LASSO})},
  journal = {IEEE Transactions on Information Theory},
  year    = {2009},
  volume  = {55},
  number  = {5},
  pages   = {2183--2202}
}

@book{wainwright2019high,
  author    = {Wainwright, M. J.},
  title     = {High-dimensional statistics: A non-asymptotic viewpoint},
  publisher = {Cambridge University Press},
  year      = {2019}
}

@article{Wang2019eEF2,
  author  = {Wang, R. X. and Xu, X. E. and Huang, L. and Chen, S. and Shao, Z. M.},
  title   = {{eEF2} kinase mediated autophagy as a potential therapeutic target for paclitaxel-resistant triple-negative breast cancer},
  journal = {Annals of Translational Medicine},
  year    = {2019},
  volume  = {7},
  number  = {23},
  pages   = {783}
}

@article{Yuan2007,
  author  = {Yuan, M. and Lin, Y.},
  title   = {Model selection and estimation in the {Gaussian} graphical model},
  journal = {Biometrika},
  year    = {2007},
  volume  = {94},
  number  = {1},
  pages   = {19--35}
}

@article{Zhang2008,
  author  = {Zhang, C. H. and Huang, J.},
  title   = {The sparsity and bias of the {LASSO} selection in high-dimensional linear regression},
  journal = {The Annals of Statistics},
  year    = {2008},
  volume  = {36},
  number  = {4},
  pages   = {1567--1594}
}

@article{Zhang2010,
  author  = {Zhang, C. H.},
  title   = {Nearly unbiased variable selection under minimax concave penalty},
  journal = {The Annals of Statistics},
  year    = {2010},
  volume  = {38},
  number  = {2},
  pages   = {894--942}
}

@article{Zhang2014,
  author  = {Zhang, C. H. and Zhang, S. S.},
  title   = {Confidence intervals for low-dimensional parameters in high-dimensional linear models},
  journal = {Journal of the Royal Statistical Society, Series B (Statistical Methodology)},
  year    = {2014},
  volume  = {76},
  number  = {1},
  pages   = {217--242}
}

@article{Zhao2006,
  author  = {Zhao, P. and Yu, B.},
  title   = {On model selection consistency of {LASSO}},
  journal = {Journal of Machine Learning Research},
  year    = {2006},
  volume  = {7},
  pages   = {2541--2563}
}

@article{Zou2005,
  author  = {Zou, H. and Hastie, T.},
  title   = {Regularization and variable selection via the elastic net},
  journal = {Journal of the Royal Statistical Society, Series B (Statistical Methodology)},
  year    = {2005},
  volume  = {67},
  number  = {2},
  pages   = {301--320}
}

@article{Zoppoli2012,
  author  = {Zoppoli, G. and Regairaz, M. and Leo, E. and Reinhold, W. C. and Varma, S. and Ballestrero, A. and Doroshow, J. H. and Pommier, Y.},
  title   = {Putative {DNA}/{RNA} helicase schlafen-11 ({SLFN11}) sensitizes cancer cells to {DNA}-damaging agents},
  journal = {Proceedings of the National Academy of Sciences of the United States of America},
  year    = {2012},
  volume  = {109},
  number  = {37},
  pages   = {15030--15035}
}

@article{Zou2006,
  author  = {Zou, H.},
  title   = {The adaptive {LASSO} and its oracle properties},
  journal = {Journal of the American Statistical Association},
  year    = {2006},
  volume  = {101},
  number  = {476},
  pages   = {1418--1429}
}

\end{document}